\documentclass[%
 amsmath,amssymb, aps,
prb
]{revtex4-2}
\usepackage{xcolor}
\usepackage{graphicx}% Include figure files
\usepackage{float}
\usepackage{dcolumn}% Align table columns on decimal point
\usepackage{bm}% bold math
\usepackage{booktabs}
\begin{document}

\preprint{APS/123-QED}

\title{Exact Solution for Non-Hermitian Free Fermions:\\ A Case Study of the XY Chain}
\author{Yuguan Li${^1}$}
\author{D. C. Liu${^2}$}
\email[]{dongchang.liu@anu.edu.au}
\author{Murray T. Batchelor${^{2,3}}$}
\email[]{murray.batchelor@anu.edu.au}

\affiliation{${^1}$Department of Fundamental and Theoretical Physics, Research School of Physics, Australian National University, Canberra ACT 2601, Australia}

\affiliation{${^2}$Mathematical Sciences Institute, Australian National University, Canberra ACT 2601, Australia}

\affiliation{${^3}$Centre for Modern Physics, Chongqing University, Chongqing 400044, The People's Republic of China}

\date{\today}

\begin{abstract}
\setlength{\emergencystretch}{1em}
We study the non-Hermitian XY spin chain with open boundary conditions and complex anisotropy. We express its quasi-energies and mode eigenvectors in a Chebyshev-polynomial representation using the quasi-energy as the spectral variable. For finite even $L$, we prove that every exceptional point is of second order. Each Chebyshev sector can contain at most one double root at fixed coupling, and the two sectors cannot become defective simultaneously. Retaining the boundary residual allows the associated generalized eigenvectors to be obtained by differentiation, giving the Jordan structure and the counting of ordinary and generalized many-body eigenstates. For the explicit $L=4$ case, the energy and eigenvector branches exhibit the characteristic square-root monodromy. The normalized left-right overlap is identified with the phase rigidity, whose vanishing corresponds to a divergent eigenvalue condition number. These results provide an exact treatment of defective eigensystems and their local spectral sensitivity within a quadratic fermionic model.
\end{abstract}

\maketitle

\tableofcontents

\section{Introduction}
\label{sec:intro}
The anisotropic XY model~\cite{LIEB1961} is a textbook example of an exactly solved quantum many-body spin chain~\cite{LIEB1961,karevski,mccoy}.
In both open and periodic boundary settings, the Jordan-Wigner transformation maps the spin chain to a quadratic spinless-fermion problem, from which much exact information about the energy spectrum and correlation functions has been obtained~\cite{LIEB1961,karevski,mccoy}.
More generally, the free-fermion construction has been extended to a free-parafermion structure which underlies a class of $Z_N$ clock models~\cite{Baxter1989,Fendley,BFreview}.
For $N>2$, the Hamiltonian of the free-parafermion $Z_N$ clock model is generically non-Hermitian with a complex energy spectrum.
In quantum mechanics, Hamiltonians are usually required to be Hermitian in order to guarantee real energy eigenvalues, unitary time evolution, and orthogonal and complete quantum states.
However, non-Hermitian Hamiltonians are also seen to be of great interest, exhibiting phenomena without any Hermitian counterparts, most notably exceptional points (EPs), at which the Hamiltonian becomes defective~\cite{Ashida2020,Heiss2012}.
This algebraic singularity plays a central role in non-Hermitian spectral and topological physics~\cite{Bergholtz2021,Ding2022}.
By extending the coupling parameter of the $Z_N$ clock model Hamiltonian to the complex plane, it has been shown that EPs occur where the quasi-energies defining the free spectrum become degenerate~\cite{Murray2023}.
A similar study of the non-Hermitian extension of the XY model obtained with a complex anisotropy parameter reveals two concentric EP rings in the auxiliary $\lambda$ plane that approach the unit circle in the infinite size limit~\cite{Murray}.
Here $\lambda=(\gamma-1)/(\gamma+1)$ in our convention.
In a subsequent study of non-Hermitian extensions of the XY chain in a magnetic field the
quasi-energies were related to topological invariants, with the EP ring seen to be located on the boundary between two distinct topological phases~\cite{Liu2025}.

Indeed, several non-Hermitian variants of the XY and related spin-chain models have been explored.
These include the investigation of geometric phase and phase diagram of the anisotropic XY chain with intrinsic rotation-time-reversal symmetry, while more recent work has examined criticality in non-Hermitian XY extensions~\cite{ZS2013RT,ZS2013GP,Liu2025}. Related dissipative or experimentally motivated non-Hermitian spin systems have also been considered~\cite{Lee2014,ZS2020}. Quantum criticality in anisotropic XY chains with staggered Dzyaloshinskii-Moriya interactions provides another related point of comparison for spin-chain critical behavior~\cite{Su2022}.
More broadly, hidden free-fermion structures in exactly solvable quantum systems have been clarified in related algebraic settings~\cite{Fendley2019FFD,Elman2021}.
Recent work has also connected non-Hermitian free-fermion critical systems to logarithmic conformal field theory~\cite{Io2026}, providing a complementary perspective on non-Hermitian free-fermion criticality.

\label{par:related-XY}
{Liu and Batchelor~\cite{Liu2025} obtain the open-chain quasi-energy spectra analytically and relate them to phase boundaries and winding numbers. Their many-body fidelity, entanglement, and dynamical calculations use exact diagonalization with periodic boundaries. Luo and Meden~\cite{LuoMeden2026} compare right--right and left--right expectation values, magnetization, and correlations in periodic non-Hermitian XY chains with a magnetic field. These results illustrate why the spectral construction and a prescription for physical expectation values must be distinguished.

\label{par:observables}
The left--right overlap below is a spectral diagnostic of self-orthogonality, distinct from the spatial biorthogonal polarization used to characterize boundary modes~\cite{Kunst2018Biorthogonal,Edvardsson2020Biorthogonal,Mandal2024Biorthogonal}. For a conditional pure state $|\psi\rangle$, an ordinary measurement of $O$ uses $\langle\psi|O|\psi\rangle/\langle\psi|\psi\rangle$, with the Hermitian-conjugate bra~\cite{Ashida2020,LuoMeden2026}. This differs from $\langle L|O|R\rangle/\langle L|R\rangle$, whose physical interpretation requires a specified state and setting.

In this paper, we study the non-Hermitian XY model with open boundary conditions. We first construct the exact free-fermion solution away from EPs and then analyze the defective regime, where a Jordan-chain construction is required. The central technical step is to express both the open-boundary quantization condition and the corresponding mode eigenvectors in the quasi-energy $\epsilon$, rather than the quasi-momentum $k$. At an EP the quasi-energy is a repeated root of the boundary polynomial. By retaining the boundary residual before differentiation, the same polynomial representation gives the associated generalized eigenvector and hence the Jordan chain. The invariant-subspace recurrence also establishes geometric multiplicity one for a repeated root within either mode family. This explicit construction of the eigenvectors and Jordan chains is the principal contribution beyond the known Jordan-Wigner reduction and open-chain quantization condition. We further examine the analytic continuation of the eigensystem around an isolated EP, obtaining the corresponding eigenvalue and eigenvector monodromy directly in the real-space open-boundary quadratic fermion problem.
Previous studies~\cite{Kawabata2019,Ding2022} identified characteristic EP phenomena, including the exchange of eigenenergies and eigenstates upon encircling an EP.
Here we show the corresponding structure directly in the open-boundary XY chain as a function of the complex anisotropy parameter.
This provides a direct demonstration, in a real-space open-boundary many-body setting, for the exchange of eigenstates upon encircling EPs.
For the explicit $L=4$ solution, the common algebraic field extension relates the energy and eigenvector branches. Their coalescence and Jordan structure distinguish an EP from a diabolic point.

In the periodic case, after resolving the fermion-parity boundary condition, the quadratic mode problem can be solved systematically by Fourier transformation~\cite{LIEB1961}.
By contrast, the open boundary XY mode matrix is not circulant, which makes an exact analytic construction more involved.
The Chebyshev-polynomial formulation developed below expresses the known open-chain recurrence in the quasi-energy variable while keeping the spectral parameter algebraic. The trigonometric \(k\)-description is recovered by setting $x(\varepsilon)=\cos2k$.

The Hamiltonian of the XY model with open boundary conditions is given by
\begin{eqnarray}
    H_{\gamma}=-\frac{1}{2}\sum_{j=1}^{L-1} \left(\frac{1+\gamma}{2}\sigma_j^x\sigma_{j+1}^x+\frac{1-\gamma}{2}\sigma_j^y\sigma_{j+1}^y\right).
\end{eqnarray}
\label{par:scope}
We use even $L=2n$ and $\gamma\in\mathbb C$.
As already remarked, the Hermitian XY model was solved exactly long ago by Lieb, Schultz and Mattis~\cite{LIEB1961} via the Jordan-Wigner transformation, which also preserves the quadratic fermionic structure for complex $\gamma$.
Both pairing terms then contain $\gamma$, without complex conjugation. Away from defective points, the canonical transformation uses dual left and right modes instead of Hermitian-conjugate creation and annihilation operators.
In other words, the exact solution of the non-Hermitian XY model can still be constructed using Lieb {\it et al.}'s solution as the starting point.
The detailed construction and subsequent analysis of the EP behaviour is presented in the following sections.

\section{Constructing fermionic operators with anticommutation relations}
\label{sec:construction}

To diagonalize the quasi-Hamiltonian of the non-Hermitian XY model, it is convenient to adopt a fermionic representation following the approach of Hamza {\it et al.}~\cite{stolz} in their treatment of the Hermitian case. 
After the Jordan-Wigner transformation, one introduces fermionic operators $c_i$ and $c_i^\dagger$ satisfying the canonical anticommutation relations. 
In this representation, the Hamiltonian is expressed in terms of the matrices
\begin{eqnarray}
    &A_{ij}=\frac{1}{2}(\delta_{i+1,j}+\delta_{i,j+1}),\quad B_{ij}=\frac{\gamma}{2} (\delta_{i+1,j}-\delta_{i,j+1}),\\
    &C=(c_1,c_2,\ldots,c_L,c_1^\dagger,c_2^\dagger,\ldots,c_L^\dagger)^T,
\end{eqnarray}
where $i,j=1,2,\ldots,L$. The quasi-Hamiltonian matrix $M$ is then defined as
\begin{eqnarray}
    M=\begin{pmatrix}
        A & B\\
        -B & -A
    \end{pmatrix}.
\end{eqnarray}

The conditions $A_{[ij]}=0$ and $B_{\{ij\}}=0$ ensure that $M$ is symmetric, $M=M^T$, so the Hamiltonian can be written in the compact form
\begin{eqnarray}
    H_{\gamma}=-\frac12 C^\dagger M C .\label{eq:nambu-normalization}
\end{eqnarray}
We use the Jordan-Wigner phase convention for which the quadratic Hamiltonian takes the form (5). It differs from the alternative convention by the staggered redefinition \(c_j\mapsto(-1)^{j-1}c_j\).
For the Hermitian case, the Hamiltonian may be diagonalized by a Bogoliubov transformation~\cite{LIEB1961,stolz}, corresponding to an orthogonal transformation $W$ satisfying
\begin{eqnarray}
    B=WC,\quad H=-\frac12B^\dagger \Lambda B .
\end{eqnarray}
In the non-Hermitian case, the matrix $M$ is no longer Hermitian. Nevertheless, its eigenvalues and eigenvectors can still be determined explicitly. The quasi-energies are given by the roots of the characteristic equation $\det(M-\varepsilon I)=0$. 
At EPs~\cite{Murray}, this polynomial develops multiple roots, signaling the degeneracy of quasi-energies and the corresponding eigenvectors. A multiple root signals a quasi-energy degeneracy. The additional defectiveness required for an EP is established below for repeated roots within either mode family.
\begin{eqnarray}
    \det(M-\varepsilon I)=\left|\begin{matrix}
        -\varepsilon I+A & B\\
        -B & -\varepsilon I-A
    \end{matrix}\right|.
\end{eqnarray}

To simplify the analysis, we introduce the matrix
\begin{eqnarray}
    S=\frac{1}{\sqrt{2}}\begin{pmatrix}
        I & I\\
        I & -I
    \end{pmatrix},
\end{eqnarray}
which will be used repeatedly below. Applying this transformation to $M-\varepsilon I$ yields
\begin{eqnarray}
    M-\varepsilon I&=&S\begin{pmatrix}
        -\varepsilon I & A-B\\
        A+B & -\varepsilon I
    \end{pmatrix}S ,\label{transform} \\
    \det(M-\varepsilon I)&=&\det\left(\varepsilon^2 I-(A+B)(A-B)\right)\label{Det} .
\end{eqnarray}

The determinant in Eq.~(\ref{Det}) coincides with that obtained in the Hermitian case~\cite{LIEB1961}, and the same structure extends directly to the quantum Ising chain. 
Thus the roots \(\{\epsilon_k\}\) of the polynomial in Eq.~(\ref{Det}) are the quasi-energies of the free fermion model, and the many-body energy spectrum is given by
%\begin{eqnarray}
%    E=\sum_j^{L/2}\pm \varepsilon_{\mathrm{I}j}\pm\varepsilon_{\mathrm{II}j}
%    \,.\label{free fermion energy}
%\end{eqnarray}
%
\begin{eqnarray}
    E=\frac12\left(\pm \varepsilon_{1}\pm\varepsilon_{2} \pm \ldots \pm \varepsilon_{L}\right)
    \,.\label{free fermion energy}
\end{eqnarray}
Here one representative is selected from each of the $L$ pairs $\{\varepsilon_k,-\varepsilon_k\}$, and the signs are independent.
Therefore, the quasi-energies $\varepsilon_k$ take the same form as in the Hermitian model. The corresponding coupled mode vectors were also obtained by Lieb {\it et al.} and satisfy
\begin{eqnarray}
    (A+B)\phi_k&=&\varepsilon_k \psi_k \label{liebeq1} \,, \\
    (A-B)\psi_k&=&\varepsilon_k \phi_k \label{liebeq2} \,.
\end{eqnarray}

It is useful to make the connection with the open-chain solution more explicit by rewriting the coupled equations (\ref{liebeq1})-(\ref{liebeq2}) in polynomial form, for even $L$ and $\gamma\neq\pm1$. The values $\gamma=\pm1$ can be treated directly in the finite matrix $M$. Define
\begin{eqnarray}
    x(\varepsilon)&=&\frac{2\varepsilon^2-1-\gamma^2}{1-\gamma^2}.
    \label{eq:xy-poly-x}
\end{eqnarray}
The point of this rewriting is that the eigenvector components can be treated as functions of \(\varepsilon\), so that the boundary quantization condition and the eigenvector components are controlled by the same algebraic variable.
Eliminating $\psi$ from the coupled equations gives a recurrence relation for the even components of $\phi$,
\begin{eqnarray}
    (4\varepsilon^2-2-2\gamma^2)\phi_{2m}
    &=&(1-\gamma^2)(\phi_{2m+2}+\phi_{2m-2}).
    \label{eq:xy-poly-even-phi-recurrence}
\end{eqnarray}
Introducing $Q_m(\varepsilon)=(1-\gamma^2)^m\phi_{2m+2}(\varepsilon)$, this becomes
\begin{eqnarray}
    Q_{m+1}(\varepsilon)&=&({4}\,\varepsilon^2-2-2\gamma^2)Q_m(\varepsilon)
    -(1-\gamma^2)^2Q_{m-1}(\varepsilon).
    \label{eq:xy-poly-Q-recurrence}
\end{eqnarray}
Hence
\begin{eqnarray}
    Q_m(\varepsilon)&=&(1-\gamma^2)^mU_m(x),
    \qquad
    U_{m+1}(x)=2xU_m(x)-U_{m-1}(x),
\end{eqnarray}
where $U_m$ is the Chebyshev polynomial of the second kind and we use the convention $U_{-1}=0$. The first family of solutions can therefore be written as
\begin{eqnarray}
    \phi^{\mathrm{I}}_{2m}(\varepsilon)&=&U_{m-1}(x(\varepsilon)),\nonumber\\
    \psi^{\mathrm{I}}_{2m+1}(\varepsilon)&=&\frac{1+\gamma}{2\varepsilon}U_m(x(\varepsilon))
    +\frac{1-\gamma}{2\varepsilon}U_{m-1}(x(\varepsilon)),
    \label{eq:xy-poly-mode-I-components}
\end{eqnarray}
with the remaining parity components set to zero. 
Setting 
\begin{equation}
(1-\gamma)/(1+\gamma)=-\lambda \,, \label{lambda}
\end{equation}
the open-boundary condition $\psi_{L+1}=0$ gives the boundary polynomial
\begin{eqnarray}
    U_{L/2}\left(x(\varepsilon^{\mathrm{I}}_k)\right)
    -\lambda \,
    U_{L/2-1}\left(x(\varepsilon^{\mathrm{I}}_k)\right)&=&0 \,.
    \label{eq:xy-poly-mode-I-boundary}
\end{eqnarray}
The second family is obtained in the same way from the even components of $\psi$ and the odd components of $\phi$:
\begin{eqnarray}
    \psi^{\mathrm{II}}_{2m}(\varepsilon)&=&U_{m-1}(x(\varepsilon)),\nonumber\\
    \phi^{\mathrm{II}}_{2m+1}(\varepsilon)&=&\frac{1-\gamma}{2\varepsilon}U_m(x(\varepsilon))
    +\frac{1+\gamma}{2\varepsilon}U_{m-1}(x(\varepsilon)),
    \label{eq:xy-poly-mode-II-components}
\end{eqnarray}
with boundary condition
\begin{eqnarray}
    U_{L/2}\left(x(\varepsilon^{\mathrm{II}}_k)\right)
    - \lambda^{-1} \,
    U_{L/2-1}\left(x(\varepsilon^{\mathrm{II}}_k)\right)&=&0 \,.
    \label{eq:xy-poly-mode-II-boundary}
\end{eqnarray}

An equivalent recurrence relation for the open-boundary XY-chain modes was given in the free-fermionic treatment by Karevski~\cite{karevski}, where it was solved in trigonometric form rather than in the Chebyshev-polynomial form derived above. 
Here the emphasis is that the recurrence relation is solved with \(x(\varepsilon)\) as the variable, so the resulting mode vectors remain directly tied to the roots of the quasi-energy characteristic polynomial.
Moreover, these two Chebyshev-polynomial equations are precisely the polynomial form of the two classes of open-chain modes in Lieb {\it et al.}'s notation. Setting $x=\cos 2k$ and using
\begin{eqnarray}
    U_n(\cos\theta)&=&\frac{\sin[(n+1)\theta]}{\sin\theta},
\end{eqnarray}
The ratios below are extended by continuity at removable singularities. Eq.~(\ref{eq:xy-poly-mode-I-boundary}) becomes
\begin{eqnarray}
    \frac{\sin(L+2)k}{\sin Lk}&=& \lambda,
    \qquad \mathrm{or} \quad
    \frac{\tan(L+1)k}{\tan k} = - \gamma,
    \label{eq:lieb-mode-I-momentum}
\end{eqnarray}
which is the quasi-momentum equation for the first kind of mode in Lieb {\it et al.}'s solution. 
Similarly, Eq.~(\ref{eq:xy-poly-mode-II-boundary}) gives
\begin{eqnarray}
    \frac{\sin(L+2)k}{\sin Lk}&=& \lambda^{-1},
    \qquad \mathrm{or} \quad
    \frac{\tan(L+1)k}{\tan k} = \gamma,
    \label{eq:lieb-mode-II-momentum}
\end{eqnarray}
corresponding to mode of the second kind. The relation $x(\varepsilon)=\cos 2k$ also yields
\begin{eqnarray}
    \varepsilon_k^2&=&\left[1-(1-\gamma^2)\sin^2 k\right],
    \qquad
    \varepsilon_k=\pm\frac{\cos k}{\cos(L+1)k}.
    \label{eq:xy-poly-lieb-quasienergy}
\end{eqnarray}

The polynomial eigenvectors given above are thus the Chebyshev-polynomial version of the trigonometric eigenvectors found by Lieb {\it et al.}~\cite{LIEB1961}, and they continue to define the non-Hermitian open chain modes for complex $\gamma$ away from EPs.
Below, this representation will be used both to label the eigenvalues and to analytically continue the corresponding eigenvectors as the roots degenerate at EPs.
Thus, up to the explicit factors $1/\varepsilon$ in Eqs.~(\ref{eq:xy-poly-mode-I-components}) and~(\ref{eq:xy-poly-mode-II-components}), the open-chain eigenvectors are encoded by polynomials in \(x(\varepsilon)\).
Multiplying the entire vector by $\varepsilon$ gives a polynomial vector in $\varepsilon$ for fixed $\gamma\neq\pm1$. At a nonzero quasi-energy this changes only the normalization; this representative is used in the Jordan-chain proof below.

Using these relations, one finds
\begin{eqnarray}
    MS\begin{pmatrix}
        \phi_k\\
        \psi_k
    \end{pmatrix}&=&S\begin{pmatrix}
        0 & A-B\\
        A+B & 0
    \end{pmatrix}\begin{pmatrix}
        \phi_k\\
        \psi_k
    \end{pmatrix} ,\\
    MS\begin{pmatrix}
        \phi_k\\
        \psi_k
    \end{pmatrix}&=&\varepsilon_k S\begin{pmatrix}
        \phi_k\\
        \psi_k
    \end{pmatrix},
\end{eqnarray}
and hence an eigenvector of $M$ may be constructed as
\begin{eqnarray}
    \frac{1}{\sqrt{2}}\begin{pmatrix}
        \phi_k+\psi_k\\
        \phi_k-\psi_k
    \end{pmatrix}.
\end{eqnarray}

Now let $V$ denote the matrix whose columns are the eigenvectors of $M$. 
Away from EPs, $M$ may be diagonalized as $M=V\Lambda V^{-1}$, where $\Lambda$ is the diagonal matrix of eigenvalues $\varepsilon_k$. 
For a Hermitian Hamiltonian, one may choose a unitary diagonalization $M=U\Lambda U^\dagger$, which preserves the standard inner-product structure. 
In the present case, $[M,M^\dagger]$ need not vanish, so a unitary transformation is not guaranteed. 
Since $M$ is symmetric, $M=M^T$, it can instead be diagonalized by an orthogonal matrix $V$, such that $M=V\Lambda V^T$.
%Consequently, the right and left eigenvectors of the Hamiltonian are not related by Hermitian conjugation.

Equivalently, after setting $x=\cos 2k$, the Chebyshev-polynomial solutions reduce to the trigonometric forms used in Lieb {\it et al.}'s solution
\begin{equation}
	\phi_{\mathrm{I}k}=A_k\begin{pmatrix}
		0\\
		\sin 2k\\
		0\\
		\sin 4k\\
		\vdots\\
		0\\
		\sin Lk
	\end{pmatrix},\quad
	\psi_{\mathrm{I}k}=-A_k \delta_k\begin{pmatrix}
		\sin Lk \\
		0\\
		\sin (L-2)k\\
		0\\
		\vdots\\
		\sin 2k\\
		0
	\end{pmatrix},
\end{equation}
subject to $k$ satisfying Eq.~(\ref{eq:lieb-mode-I-momentum}). Also  
\begin{equation}
	\phi_{\mathrm{II}k}=A_k\begin{pmatrix}
		\sin Lk \\
		0\\
		\sin (L-2)k\\
		0\\
		\vdots\\
		\sin 2k\\
		0
	\end{pmatrix},\quad
	\psi_{\mathrm{II}k}=-A_k\delta_k\begin{pmatrix}
		0\\
		\sin 2k\\
		0\\
		\sin 4k\\
		\vdots\\
		0\\
		\sin Lk
	\end{pmatrix},
\end{equation}
subject to $k$ satisfying Eq.~(\ref{eq:lieb-mode-II-momentum}). 
Here $A_k$ is the normalization constant fixed by $2A_k^2\sum_{m=1}^{L/2}\sin^2(2mk)=1$ and \(\delta_k=\pm 1\) is chosen by the sign of \(\pm\varepsilon_k\) which are the corresponding eigenvalues.
This uses the transpose rather than Hermitian-conjugate bilinear form. The sine vector must be nonzero. At a removable singularity one uses the polynomial representative or its nonzero limit. At a degeneracy with a complete eigenspace, a complex-orthogonal basis is chosen within that eigenspace. The vanishing bilinear norm at an EP prevents this normalization. 

The matrix $V$ can be written as
\begin{eqnarray}
    V=\frac{1}{\sqrt{2}}\begin{pmatrix}
        I & I \\
        I & -I
    \end{pmatrix}\begin{pmatrix}
        \phi_{\mathrm{I}} & \phi_{\mathrm{II}}\\
        \psi_{\mathrm{I}} & \psi_{\mathrm{II}}
    \end{pmatrix}.
\end{eqnarray}
Since $M$ is a symmetric matrix and its eigenvectors may be chosen orthogonal, the inverse of $V$ takes the form
\begin{eqnarray}
	V^{-1}=\frac{1}{\sqrt{2}}\begin{pmatrix}
		\phi_{\mathrm{I}}^{T} & \psi_{\mathrm{I}}^{T}\\
		\phi_{\mathrm{II}}^{T} & \psi_{\mathrm{II}}^{T}
	\end{pmatrix}\begin{pmatrix}
		I & I\\
		I & -I
	\end{pmatrix}.
\end{eqnarray}
The {corresponding orthogonality relations} are
\begin{eqnarray}
    \sum_{\mu=1}^{L}\phi_{\sigma i}^{\mu}\phi_{\sigma' j}^{\mu}+\psi_{\sigma i}^{\mu}\psi_{\sigma' j}^{\mu}=\delta_{ij}\delta_{\sigma\sigma'} \,,\label{orth}\\
    \sum_{k=1}^{L}\varphi_{\mathrm{I} k}^{\mu}\xi_{\mathrm{I} k}^{ \nu}+\varphi_{\mathrm{II} k}^{\mu}\xi_{\mathrm{II} k}^{ \nu}=\delta_{\mu\nu}\delta_{\varphi\xi} \,.
\end{eqnarray}
Here \(\varphi\) and \(\xi\) can be chosen as \(\phi\) or \(\psi\).
To specify the inverse superscripts used below, write
\begin{equation}
 P=S V=\begin{pmatrix}\phi_{\mathrm I}&\phi_{\mathrm {II}}\\
 \psi_{\mathrm I}&\psi_{\mathrm {II}}\end{pmatrix},\qquad
 P^{-1}=\begin{pmatrix}\phi_{\mathrm I}^{-1}&\psi_{\mathrm I}^{-1}\\
 \phi_{\mathrm {II}}^{-1}&\psi_{\mathrm {II}}^{-1}\end{pmatrix}.
 \label{eq:inverse-blocks}
\end{equation}
These superscripts label blocks of the inverse transformation. They are not componentwise reciprocals of vectors. Away from EPs, $P^{-1}=P^T$. For a Jordan basis the inverse must instead be calculated from the full matrix containing generalized eigenvectors.

We then define biorthogonal fermionic operators by
\begin{eqnarray}
    R=V^{-1}C=(R,R^*),\qquad L^*=C^{\dagger}V=(L^*,L),
\end{eqnarray}
which gives
\begin{eqnarray}
    R_k&=&\frac{1}{\sqrt{2}}\left[\phi_{\mathrm{I}k}^{\mu}(c_\mu+c_\mu^\dagger)+\psi_{\mathrm{I}k}^{\mu}(c_\mu-c_\mu^\dagger)\right],\\
    R_k^*&=&\frac{1}{\sqrt{2}}\left[\phi_{\mathrm{II}k}^{\mu}(c_\mu+c_\mu^\dagger)+\psi_{\mathrm{II}k}^{\mu}(c_\mu-c_\mu^\dagger)\right],\\
    L^*_k&=&\frac{1}{\sqrt{2}}\left[\phi_{\mathrm{I}k}^{\mu}(c_\mu+c_\mu^\dagger)-\psi_{\mathrm{I}k}^{\mu}(c_\mu-c_\mu^\dagger)\right],\\
    L_k&=&\frac{1}{\sqrt{2}}\left[\phi_{\mathrm{II}k}^{\mu}(c_\mu+c_\mu^\dagger)-\psi_{\mathrm{II}k}^{\mu}(c_\mu-c_\mu^\dagger)\right].
\end{eqnarray}

Using the orthogonality relations~(\ref{orth}) together with the explicit forms of $\phi_k$ and $\psi_k$, these operators satisfy the following anticommutation relations, with details given in Appendix~A:
\begin{eqnarray}
    &\{L^*_i, R_j\}=\{L_i, R^*_j\}=\delta_{ij}I,\\
    &\{L^*_i,L_j\}=\{L^*_i,R^*_j\}=\{R_i,R^*_j\}=\{L_i,R_j\}=0,\\
    &R_iR_i=R_i^* R_i^*=0.
\end{eqnarray}
In terms of these operators, the Hamiltonian takes the diagonal form
\begin{eqnarray}
	H_{\gamma}=-\frac12C^\dagger M C=-\frac12\sum_{k=1}^{L}\left(L^*_k \varepsilon_{\mathrm{I}k} R_k + L_k \varepsilon_{\mathrm{II}k} R_k^*\right).
\end{eqnarray}
The sum runs over all signed eigenvalues within each mode family.

From the characteristic equation~(\ref{Det}), the spectrum appears in pairs, such that if $\varepsilon_k$ is an eigenvalue of $M$, then $-\varepsilon_k$ is also an eigenvalue. By ordering the eigenvectors in $V$ appropriately, the diagonal matrix $\Lambda$ may be written as
\begin{eqnarray}
	\Lambda=\text{diag}(\varepsilon_{\mathrm{I}1},-\varepsilon_{\mathrm{I}1},\dots,\varepsilon_{\mathrm{I}L/2},-\varepsilon_{\mathrm{I}L/2},\varepsilon_{\mathrm{II}1},-\varepsilon_{\mathrm{II}1},\dots,\varepsilon_{\mathrm{II}L/2},-\varepsilon_{\mathrm{II}L/2}).
\end{eqnarray}
Recall that the quasi-energies are given in Eq.~(\ref{eq:xy-poly-lieb-quasienergy}), where $L$ is even. 
%\begin{eqnarray}
%    &\varepsilon_k^2=\left[1-(1-\gamma^2)\sin^2 k\right],
%\end{eqnarray}
%
Introduce $T=\left(\begin{smallmatrix}0&I\\I&0\end{smallmatrix}\right)$, so that $TMT=-M$ and $C^\dagger=C^TT$. For a normalized column $v_{\sigma,+k}$ of $V$, choose its opposite-energy partner as $v_{\sigma,-k}=T v_{\sigma,+k}$. This fixes the otherwise arbitrary relative sign. The independent canonical fermionic operators are then
\begin{eqnarray}
 f_{\mathrm I k}=R_{2k}=L_{2k-1}^*,&&
 \bar f_{\mathrm I k}=L_{2k}^*=R_{2k-1},\nonumber\\
 f_{\mathrm {II}k}=R_{2k}^*=L_{2k-1},&&
 \bar f_{\mathrm {II}k}=L_{2k}=R_{2k-1}^*.
 \label{eq:canonical-modes}
\end{eqnarray}
They satisfy $\{f_{\sigma k},\bar f_{\sigma'j}\}=\delta_{\sigma\sigma'}\delta_{kj}$ and $\{f_{\sigma k},f_{\sigma'j}\}=\{\bar f_{\sigma k},\bar f_{\sigma'j}\}=0$, as proved in Appendix~A. The bar is a label for the canonical partner and does not denote Hermitian conjugation. Thus
\begin{equation}
 H_\gamma=\sum_{\sigma=\mathrm I,\mathrm {II}}\sum_{k=1}^{L/2}
 \varepsilon_{\sigma k}\left(\bar f_{\sigma k}f_{\sigma k}-\frac12\right).
 \label{energy}
\end{equation}

We choose a vacuum state $|\Omega\rangle$ such that
\begin{equation}
 f_{\sigma k}|\Omega\rangle=0.\label{Vacuum}
\end{equation}
The vacuum energy is then defined as
\begin{equation}
 E_0=-\frac12\sum_{\sigma=\mathrm I,\mathrm {II}}\sum_{k=1}^{L/2}\varepsilon_{\sigma k}.
\end{equation}
This expression guarantees the structure of the free fermion (\ref{free fermion energy}).
The full set of right eigenvectors is then given by
\begin{equation}
 |\Psi_\alpha\rangle=\prod_{\sigma,k}(\bar f_{\sigma k})^{\alpha_{\sigma k}}|\Omega\rangle,\qquad\alpha_{\sigma k}=0,1,
\end{equation}
with any fixed ordering of the product. Their energies are $E_0+\sum_{\sigma,k}\alpha_{\sigma k}\varepsilon_{\sigma k}$. A dual vacuum and the $f_{\sigma k}$ construct the corresponding left basis. 
Together these form a biorthogonal basis. 

The exact energy eigenvalues and eigenvectors for the \(L=4\) open chain are given in Appendix B for the purposes of comparison and testing.
The $L=4$ example makes explicit how the above polynomial construction reproduces the full energy spectrum and the corresponding eigenvectors. 
This is also beneficial for the discussion of EPs given in the following section.

\section{Exceptional points}
\label{sec:EP}

A multiple root of the characteristic polynomial $\det(M-\varepsilon I)$ signals a quasi-energy degeneracy. For the degeneracies considered below, their exceptional-point character is established by the coalescence of the corresponding eigenvectors and the explicit Jordan-chain construction. 
The Chebyshev-polynomial formulation makes this multiple-root structure explicit at the level of the open-boundary quantization conditions, because the same quasi-energy variable \(\varepsilon\) appears both in the boundary polynomials and in the eigenvector components.
In previous work~\cite{Murray}, EPs were identified by differentiating Eq.~(\ref{eq:lieb-mode-I-momentum}) with respect to $k$ and imposing that both the equation and its derivative vanish simultaneously. 
This yields the condition
\begin{eqnarray}
    (L+2)\sin Lk_{EP}\cos(L+2)k_{EP}-L\sin(L+2)k_{EP}\cos Lk_{EP}=0 .
\end{eqnarray}
Since \(dx/dk=-2\sin 2k\), this condition also holds at a simple root when \(\sin 2k=0\). The corresponding endpoints \(x=\pm1\) are excluded by Eq.~(\ref{eq:endpoint-derivative}) and parity.

The eigenvectors collapse at EPs~\cite{Heiss2012,Bergholtz2021}. 
In terms of the eigenvector matrix $V$ defined in the previous section, this means that some columns of $V$ become linearly dependent, so that $V$ is not invertible and the quasi-Hamiltonian matrix $M$ becomes defective~\cite{Heiss2012}. 

The polynomial formulation also determines their geometric multiplicity. Define
\begin{equation}
 P_{\mathrm I}(\varepsilon,\gamma)=U_n(x)-\lambda U_{n-1}(x),\qquad
 P_{\mathrm {II}}(\varepsilon,\gamma)=U_n(x)-\lambda^{-1}U_{n-1}(x).
\end{equation}
Set $a=(1+\gamma)/2$ and $b=(1-\gamma)/2$. The matrix $SMS$ preserves the two subspaces
\begin{equation}
 \mathcal U_{\mathrm I}=\{(\phi,\psi):\phi_{\rm odd}=\psi_{\rm even}=0\},\qquad
 \mathcal U_{\mathrm {II}}=\{(\phi,\psi):\phi_{\rm even}=\psi_{\rm odd}=0\}.
\end{equation}
In the site ordering $(\psi_1,\phi_2,\psi_3,\ldots,\phi_L)$, the first restriction $K_{\mathrm I}$ is tridiagonal with successive off-diagonal entries $a,b,a,\ldots,a$. The second restriction $K_{\mathrm {II}}$ has $b,a,b,\ldots,b$. Consequently,
\begin{equation}
 \det(\varepsilon I-K_\sigma)=(ab)^n P_\sigma(\varepsilon,\gamma),\qquad
 \sigma=\mathrm I,\mathrm {II}.
 \label{eq:sector-characteristic}
\end{equation}
When $\gamma\neq\pm1$, all of these off-diagonal entries are nonzero. The first component of an eigenvector then determines every other component through the tridiagonal recurrence. If it vanishes, the entire vector vanishes. Thus each eigenvalue of either restriction has geometric multiplicity one. A repeated root within one $P_\sigma$ is therefore defective. A coincidence of roots belonging to different invariant subspaces does not by itself establish an EP. For example, the two families coincide at the Hermitian point $\gamma=0$.

A complete basis at the EP therefore requires the Jordan normal form and the associated generalized eigenvectors. 
Since the quasi-Hamiltonian matrix $M$ is defective at EPs, it must be written in Jordan form, and the problem reduces to constructing the generalized eigenvectors that complete the Jordan chain.
Related finite-dimensional descriptions of passage through EPs based on Jordan blocks and transition matrices have been developed in quasi-Hermitian models~\cite{Znojil2020Passage}. The construction below plays an analogous role for the free-fermion quasi-Hamiltonian of the open XY chain.

We now construct the generalized eigenvectors, retaining the final boundary equation throughout. From the unnormalized vectors in Eqs.~(\ref{eq:xy-poly-mode-I-components}) and (\ref{eq:xy-poly-mode-II-components}), choose the polynomial representative
\begin{equation}
 w_\sigma(\varepsilon,\gamma)=\varepsilon S\binom{\phi^\sigma(\varepsilon,\gamma)}{\psi^\sigma(\varepsilon,\gamma)},\qquad
 q_{\mathrm I}=S\binom{e_L}{0},\quad q_{\mathrm {II}}=S\binom{0}{e_L},
\end{equation}
where $e_L$ is the last unit vector in $\mathbb C^L$. Direct substitution into all $2L$ equations gives the exact residual
\begin{equation}
 [M(\gamma)-\varepsilon I]w_\sigma(\varepsilon,\gamma)
 =-abP_\sigma(\varepsilon,\gamma)q_\sigma.
 \label{eq:boundary-residual}
\end{equation}
Before multiplication by $\varepsilon$, the omitted boundary value in mode $\mathrm I$ is $\psi_{L+1}=aP_{\mathrm I}/\varepsilon$, giving the residual $-b\psi_{L+1}$ in the last $\phi$ equation. In mode $\mathrm {II}$, it is $\phi_{L+1}=bP_{\mathrm {II}}/\varepsilon$, giving $-a\phi_{L+1}$ in the last $\psi$ equation. Multiplication of the full vector by $\varepsilon$ gives Eq.~(\ref{eq:boundary-residual}), and all other components vanish identically. This verifies the complete polynomial identity, including the boundary equation, before differentiating.

For a double root $\varepsilon_\star\neq0$ at fixed $\gamma=\gamma_{EP}\neq\pm1$, require
\begin{equation}
 P_\sigma(\varepsilon_\star,\gamma_{EP})=
 \partial_\varepsilon P_\sigma(\varepsilon_\star,\gamma_{EP})=0,
 \qquad \partial_\varepsilon^2P_\sigma(\varepsilon_\star,\gamma_{EP})\neq0.
 \label{eq:double-root-condition}
\end{equation}
The vector $w_\sigma$ is polynomial in $\varepsilon$ and is nonzero at this root because its seed component in the $S$ basis equals $\varepsilon_\star$. Differentiating Eq.~(\ref{eq:boundary-residual}) at fixed $\gamma$ therefore gives
\begin{equation}
 (M-\varepsilon_\star I)u_0=0,\qquad
 (M-\varepsilon_\star I)u_1=u_0,\qquad
 u_0=w_\sigma(\varepsilon_\star,\gamma_{EP}),\quad
 u_1=\left.\partial_\varepsilon w_\sigma\right|_{\varepsilon_\star,\gamma_{EP}}.
 \label{eq:derivative-jordan-chain}
\end{equation}
The two vectors are linearly independent, and Eq.~(\ref{eq:sector-characteristic}) ensures that their Jordan chain has exactly length two. One may add a multiple of $u_0$ to $u_1$. The zero parity components of the recurrence vectors vanish identically as functions of $\varepsilon$, so the differentiated vectors stay in the same invariant subspace.

It remains to determine whether higher-order quasiparticle EPs, or simultaneous defects in the two mode families, can occur.  For the present uniform zero-field chain both possibilities can be excluded exactly.
To show this, write \(n=L/2\) and either boundary polynomial as
\begin{equation}
P_n(x,\zeta)=U_n(x)-\zeta U_{n-1}(x),
\qquad
\zeta=\lambda\ \hbox{or}\ \lambda^{-1}.
\end{equation}
The Chebyshev recurrence and differential identities give
\begin{equation}
(1-x^2)\partial_x P_n
={}n(\zeta-x)P_n
+\left[n+1+n\zeta^2-(2n+1)\zeta x\right]U_{n-1}(x).
\end{equation}
Since consecutive Chebyshev polynomials have no common zero, a repeated root with \(\zeta\neq0\) must satisfy
\begin{equation}
x_*=\frac{n+1+n\zeta^2}{(2n+1)\zeta}.
\end{equation}
Thus there is at most one repeated root in either mode family at a fixed coupling.  The endpoints \(x=\pm1\) cannot be multiple roots.  For example, a root at \(x=1\) requires \(\zeta=(n+1)/n\), for which
\begin{equation}
\partial_x P_n\left(1,\frac{n+1}{n}\right)
=\frac{(n+1)(2n+1)}{3}\neq0,
\label{eq:endpoint-derivative}
\end{equation}
and the result at \(x=-1\) follows by parity.  At an interior repeated root, differentiating once more gives
\[
(1-x_*^2)P_n''(x_*,\zeta)
=-(2n+1)\zeta U_{n-1}(x_*)\neq0.
\]
Hence every multiple root of either boundary polynomial is exactly double.

No additional multiplicity is introduced by the dependence of \(x\) on \(\epsilon^2\).  For finite nonzero \(\lambda\), setting \(\epsilon=0\) gives
\begin{equation}
x=\frac{\lambda+\lambda^{-1}}{2},
\qquad
P_I=\lambda^{-n},
\qquad
P_{II}=\lambda^n,
\end{equation}
so zero quasienergy never satisfies either boundary condition.  Since
\begin{equation}
\frac{\partial x}{\partial\epsilon}
=\frac{4\epsilon}{1-\gamma^2},
\end{equation}
the multiplicity of a nonzero root in \(x\) and in \(\epsilon\) is the same.  Thus every defective quasiparticle eigenvalue considered here is of order two. 
The excluded values \(\gamma=\pm1\) are Hermitian Ising limits and hence cannot be exceptional points.

The two mode families also cannot become defective simultaneously.  Up to its leading coefficient, \(P_n(x,\zeta)\) is the characteristic polynomial of the tridiagonal matrix
\begin{equation}
T_n(\zeta)=\frac12
\begin{pmatrix}
0&1&&\\
1&0&\ddots&\\
&\ddots&\ddots&1\\
&&1&\zeta
\end{pmatrix}.
\end{equation}
At a defective eigenvalue \(x_*\), let \(v\) be the corresponding eigenvector.  Taking the imaginary part of \(v^\dagger T_n(\zeta)v=x_*v^\dagger v\) gives
\begin{equation}
\frac{\operatorname{Im}x_*}{\operatorname{Im}\zeta}
=
\frac{|v_n|^2}{2\|v\|^2}>0,
\end{equation}
where \(\operatorname{Im}\zeta\neq0\) because \(T_n(\zeta)\) is Hermitian for real \(\zeta\), and \(v_n\neq0\) by irreducibility.  On the other hand, the expression for \(x_*\) gives
\begin{equation}
\frac{\operatorname{Im}x_*}{\operatorname{Im}\zeta}
=
\frac{1}{2n+1}
\left(n-\frac{n+1}{|\zeta|^2}\right).
\end{equation}
Consequently every defective sector satisfies
\begin{equation}
|\zeta|^2>\frac{n+1}{n}.
\end{equation}
The two physical values \(\zeta=\lambda\) and \(\zeta=\lambda^{-1}\) therefore cannot both satisfy the defect condition.  Equivalently,
\begin{equation}
|\lambda|>\sqrt{\frac{n+1}{n}}
\end{equation}
for an EP in one family, whereas an EP in the reciprocal family requires
\begin{equation}
|\lambda|<\sqrt{\frac{n}{n+1}}.
\end{equation}
This gives an exact finite-size exclusion annulus containing \(|\lambda|=1\), whose width tends to zero as \(L\) increases.
At a common root \(x\) of the two boundary polynomials, subtraction gives \((\lambda-\lambda^{-1})U_{n-1}(x)=0\). Since \(U_{n-1}(x)\ne0\) at such a root, this requires \(\lambda^2=1\), which for finite \(\gamma\) occurs only at the Hermitian point \(\gamma=0\). Thus each defective eigenvalue has algebraic multiplicity two in the full matrix \(M\).

It follows that at every exceptional coupling of the finite even chain there is one coalescing quasiparticle pair in a single mode family.  Particle-hole symmetry produces the corresponding length-two blocks at \(+\epsilon_{\rm EP}\) and \(-\epsilon_{\rm EP}\), while all remaining quasiparticle modes are diagonalizable.

The Jordan form of $M$ consists of those two blocks and $2L-4$ blocks of size one. 
At an EP, the quasi-Hamiltonian matrix admits the Jordan decomposition
\begin{eqnarray}
    M&=&V_{EP}JV_{EP}^{-1} \,,\\
    J&=&\bigoplus_{k=1}^{2L-2}J_k \,.
\end{eqnarray}
Here $J$ is the Jordan normal form of $M$ and $J_k$ are Jordan blocks. 
In particular, there are two non-trivial Jordan blocks $J_{EP_{\pm}}$ with eigenvalues $\pm \varepsilon_{EP}$ and unit off-diagonal entries. 
The columns of $V_{EP}$ are generalized eigenvectors of the matrix $M$, and we choose the form
\begin{eqnarray}
    V_{EP}=S\begin{pmatrix}
        \phi_{\mathrm{I}} &\cdots&\phi_{EP_+}&\phi_{EP_+}^{\text{ker}}&\phi_{EP_-}&\phi_{EP_-}^{\text{ker}}&\cdots& \phi_{\mathrm{II}}\\
        \psi_{\mathrm{I}} &\cdots&\psi_{EP_+}&\psi_{EP_+}^{\text{ker}}&\psi_{EP_-}&\psi_{EP_-}^{\text{ker}}&\cdots& \psi_{\mathrm{II}}
    \end{pmatrix},
\end{eqnarray}
where $S$ is the matrix defined in the previous section. 
For each defective eigenvalue, the corresponding eigenvector and generalized-eigenvector columns may be chosen as the vectors $u_0$ and $u_1$ constructed above.

The generalized eigenvectors at the EP satisfy
\begin{eqnarray}
    (M\mp\varepsilon_{EP}I)S\begin{pmatrix}
        \phi_{EP_{\pm}}^{\text{ker}}\\
        \psi_{EP_{\pm}}^{\text{ker}}
    \end{pmatrix}=S\begin{pmatrix}
        \phi_{EP_{\pm}}\\
        \psi_{EP_{\pm}}
    \end{pmatrix}.
\end{eqnarray}
Using the block form introduced in the previous section, this implies
\begin{eqnarray}
    \begin{pmatrix}
        \mp \varepsilon_{EP} I & A-B\\
        A+B & \mp \varepsilon_{EP} I
    \end{pmatrix}\begin{pmatrix}
        \phi_{EP_{\pm}}^{\text{ker}}\\
        \psi_{EP_{\pm}}^{\text{ker}}
    \end{pmatrix}&=&\begin{pmatrix}
        \phi_{EP_{\pm}}\\
        \psi_{EP_{\pm}}
    \end{pmatrix}.
\end{eqnarray}
In components, using the explicit forms of $A$ and $B$, this gives
\begin{eqnarray}
    &&\phi_{EP_{\pm},i}=\mp \varepsilon_{EP} \phi_{EP_{\pm},i}^{\text{ker}} + \frac{1-\gamma}{2}\psi_{EP_{\pm},i+1}^{\text{ker}} + \frac{1+\gamma}{2}\psi_{EP_{\pm},i-1}^{\text{ker}} \,,\\
    &&\psi_{EP_{\pm},i}=\mp \varepsilon_{EP} \psi_{EP_{\pm},i}^{\text{ker}} + \frac{1+\gamma}{2}\phi_{EP_{\pm},i+1}^{\text{ker}} + \frac{1-\gamma}{2}\phi_{EP_{\pm},i-1}^{\text{ker}}\label{kereq} \,.
\end{eqnarray}

The full matrix $V_{EP}$ is invertible and gives
\begin{equation}
 H_\gamma=-\frac12\,\mathcal L_{EP}J\mathcal R_{EP},\qquad
 \mathcal R_{EP}=V_{EP}^{-1}C,\quad \mathcal L_{EP}=C^\dagger V_{EP}.
\end{equation}
Its inverse defines the dual Jordan basis. In particular, $V_{EP}^{-1}$ cannot be replaced by $V_{EP}^T$. The symbols $\phi^{-1}$ and $\psi^{-1}$ continue to denote its inverse-transformation blocks as defined in the previous section.

The independent canonical operators in the defective sector can be constructed explicitly. Choose the chain at $+\varepsilon_{EP}$ to obey
\begin{equation}
 u_0^Tu_0=u_1^Tu_1=0,\qquad u_0^Tu_1=1.
 \label{eq:Jordan-bilinear-normalization}
\end{equation}
This is possible by a common rescaling of the chain and a shift $u_1\mapsto u_1+\beta u_0$: the eigenvalue $+\varepsilon_{EP}$ has algebraic multiplicity two, so the bilinear form on its generalized eigenspace is nondegenerate and $u_0^Tu_1\neq0$. Define
\begin{equation}
 \bar f_1=u_0^TC,\quad \bar f_2=u_1^TC,\qquad
 f_1=C^\dagger u_1,\quad f_2=C^\dagger u_0.
 \label{eq:Jordan-canonical-modes}
\end{equation}
These are linear combinations of the $R/L$ operators obtained from $V_{EP}$ and its inverse. Appendix~A proves their canonical anticommutation relations. Together with the unaffected canonical modes they give
\begin{equation}
 H_\gamma=\sum_{(\sigma,k)\notin EP}\varepsilon_k\left(\bar f_kf_k-\frac12\right)
 +\varepsilon_{EP}(\bar f_1f_1+\bar f_2f_2-1)+\bar f_1f_2.
 \label{Ham_ep}
\end{equation}
The unit coefficient of the last term is fixed by the normalization of the Jordan chain in Eq.~(\ref{eq:derivative-jordan-chain}).

Let $|00\rangle$ be annihilated by $f_1,f_2$, and set $|10\rangle=\bar f_1|00\rangle$, $|01\rangle=\bar f_2|00\rangle$, and $|11\rangle=\bar f_1\bar f_2|00\rangle$. The restriction $H_{EP}$ to these two modes acts as
\begin{equation}
 H_{EP}|00\rangle=-\varepsilon_{EP}|00\rangle,\quad
 H_{EP}|11\rangle=\varepsilon_{EP}|11\rangle,\quad
 H_{EP}|10\rangle=0,\quad H_{EP}|01\rangle=|10\rangle.
\end{equation}
This space has dimension four and contains three independent eigenvectors and one generalized eigenvector. Each of the $2^{L-2}$ configurations of the unaffected modes contributes such a space, so
\begin{equation}
 N_{EP}=3\times2^{L-2},\qquad
 N_{\rm generalized}=2^{L-2}.
 \label{eq:state-count}
\end{equation}
The many-body construction does not increase the Jordan order.  In the defective two-mode sector the nilpotent part is
\[
N=\bar f_1f_2,
\qquad
N^2=0.
\]
Its Jordan form is therefore
\[
J_1(-\epsilon_{\rm EP})
\oplus J_2(0)
\oplus J_1(\epsilon_{\rm EP}).
\]
All remaining fermionic modes are diagonalizable spectators.  Their \(2^{L-2}\) occupation configurations reproduce this four-dimensional factor with different energy shifts but cannot enlarge its nilpotent index.  Hence at an exceptional coupling the full \(2^L\)-dimensional spin Hamiltonian contains
\[
2^{L-2}\quad\hbox{blocks of size two},
\qquad
2^{L-1}\quad\hbox{blocks of size one}.
\]
Thus every exceptional point of the uniform zero-field open chain with finite even \(L\) is an EP2, despite its extensive replication through the many-body spectrum.

By choosing the real part of the quasi-energies to be positive, we define the ground state as the state with the lowest real part of the energy spectrum. 
Here all selected real parts are assumed strictly positive, including $\operatorname{Re}\varepsilon_{EP}>0$.
The ground state energy is given by
\begin{equation}
 E_0=-\frac12\sum_{(\sigma,k)\notin EP}\varepsilon_k-\varepsilon_{EP}
\end{equation}
The vacuum is a nondegenerate eigenstate. The defective one-particle level is the first level ordered by real part only when $\operatorname{Re}\varepsilon_{EP}$ is the smallest positive excitation real part.

The restriction to even $L$ is essential. For \(L=3\), for example, at \(\gamma=-i\) one finds \(H_\gamma^3=0\), with \(\operatorname{rank}H_\gamma=4\) and \(\operatorname{rank}H_\gamma^2=2\), giving the Jordan form \(J_3(0)\oplus J_3(0)\oplus J_1(0)\oplus J_1(0)\). Thus the EP2 bound is a consequence of the even-chain Chebyshev-sector structure rather than of free-fermion solvability alone.

Previous work identified the EP rings of the complex-anisotropy XY model in the $\lambda$ plane~\cite{Murray}. 
For $L=4$, EPs are found precisely at $\gamma=0.6\pm0.8 \,\mathrm{i}$ and $\gamma=-0.6\pm0.8 \,\mathrm{i}$. 
For larger $L$, the EP locations are obtained by imposing Eq.~(\ref{eq:double-root-condition}) in either mode family. 
These are tabulated in Appendix C up to $L=14$.
A standard characterization of an EP is the degeneracy and self-orthogonality of the corresponding left and right eigenvectors, so that the biorthogonal overlap vanishes, $\langle L|R\rangle=0$, at the EP~\cite{Heiss2012,Ding2022}. 
For the complex-symmetric Hamiltonian considered here, if
\(H_\gamma v=Ev\) and \((H_\gamma-EI)w=v\), then
\(v^Tv=v^T(H_\gamma-EI)w=0\).
To display this property, we use a different normalization from the biorthogonal normalization of the fermionic operators. For a nonzero, unnormalized right eigenvector \(v\), set
\begin{equation}
    |R\rangle=\frac{v}{\sqrt{v^\dagger v}},\qquad
    \langle L|=\frac{v^T}{\sqrt{v^\dagger v}},\qquad
    \rho(\gamma)=\langle L|R\rangle
    =\frac{v^Tv}{v^\dagger v}.
    \label{eq:overlap-normalization}
\end{equation}
Both the right ket and the ket corresponding to the left bra have unit Euclidean norm. Their mutual overlap is not fixed to unity. The modulus \(|\rho|\), often called the phase rigidity, is independent of the nonzero scalar multiplying \(v\) and vanishes for the coalesced eigenvector. By contrast, under \(v\mapsto a v\), the complex quantity transforms as \(\rho\mapsto(a/a^*)\rho\), so its real and imaginary parts require a phase convention.

For the plots we use the explicit vectors \(v^{(5)}_{+,s}\) in Appendix~\ref{app:L4}, whose first component is fixed to \(1\) before Euclidean normalization. The square roots in their energies are taken on their principal branches, with the two choices of \(s\) distinguished explicitly below. Figure~\ref{figa} compares the resulting \(|\rho|\) with exact diagonalization along \(\Re\gamma=0.6\). Figure~\ref{figb} extends the calculation to a neighborhood of \(\gamma_{\rm EP}=0.6+0.8\mathrm{i}\). For this pair the two branches have the same modulus \(|\rho|\). The notation \(\langle L|R\rangle\) on the axis of Figure~\ref{figa} refers to Eq.~\eqref{eq:overlap-normalization}.

The overlap is used here as a scalar spectral diagnostic. The coalescence of eigenvectors is established separately by the exact solution and the Jordan chains. It is also distinct from an expectation value in conditional non-unitary evolution, where the normalized physical-state expectation is \(\langle\psi(t)|O|\psi(t)\rangle/\langle\psi(t)|\psi(t)\rangle\), with the ordinary Hermitian-conjugate bra. Nor is this scalar overlap the biorthogonal polarization, which involves spatial information and is used to diagnose boundary-state topology and bulk--boundary correspondence~\cite{Kunst2018Biorthogonal,Edvardsson2020Biorthogonal,Mandal2024Biorthogonal}.

\begin{figure}[ht]
        \centering
        \includegraphics[width=0.7\textwidth]{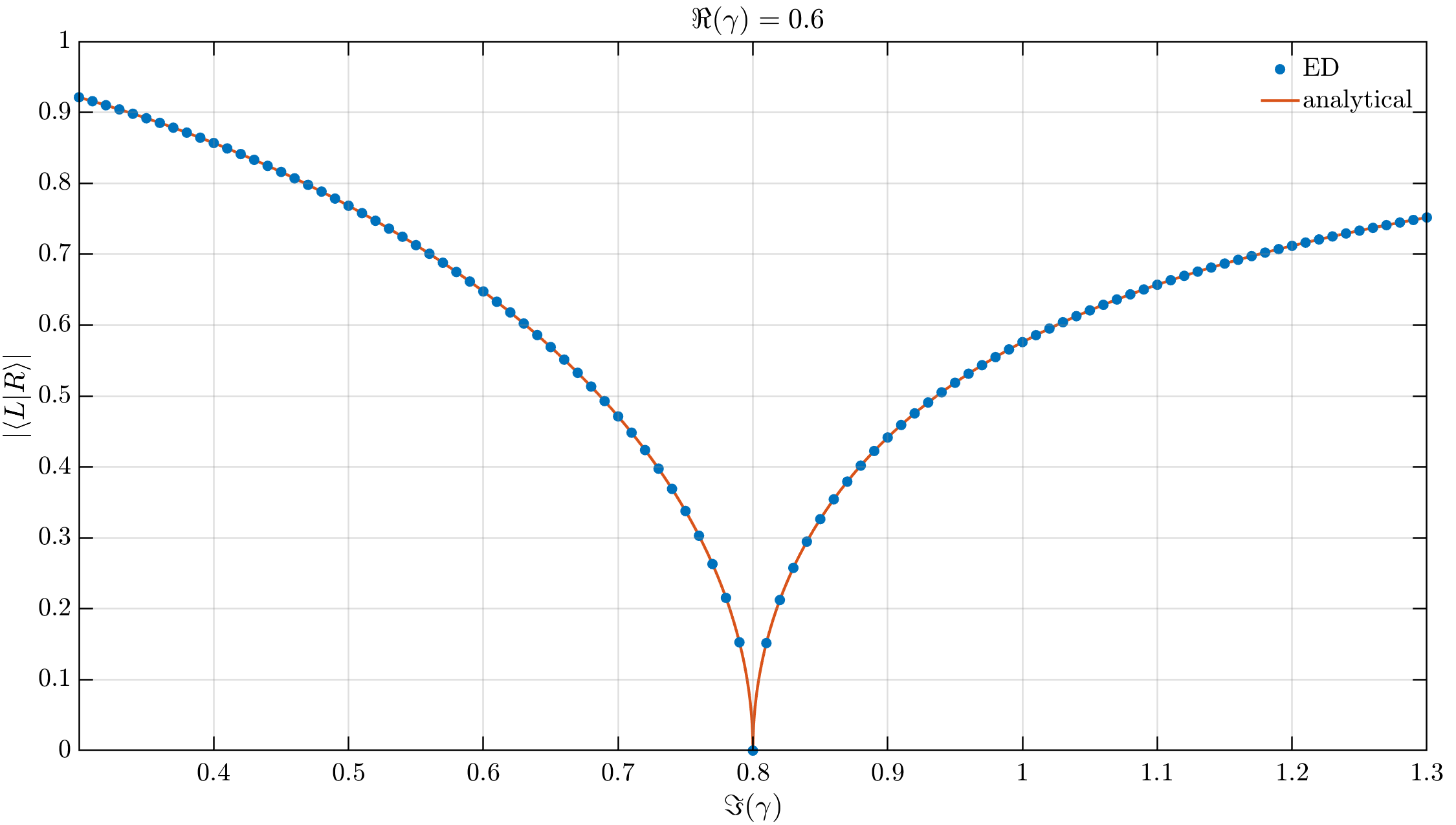}
        \caption{Plot of the inner product $|\langle L|R\rangle|$ as a function of the complex parameter $\gamma$ for $L=4$. Lines are calculated using the procedure described in the text, and dots away from the EP are obtained by exact diagonalization. At $\gamma=0.6+0.8\,\mathrm{i}$, the plotted zero marker denotes the exact analytical limit, not an ED result. The normalization is given in Eq.~\eqref{eq:overlap-normalization}.}
        \label{figa}
\end{figure}

\begin{figure}[ht]
        \centering
        \includegraphics[width=0.9\textwidth]{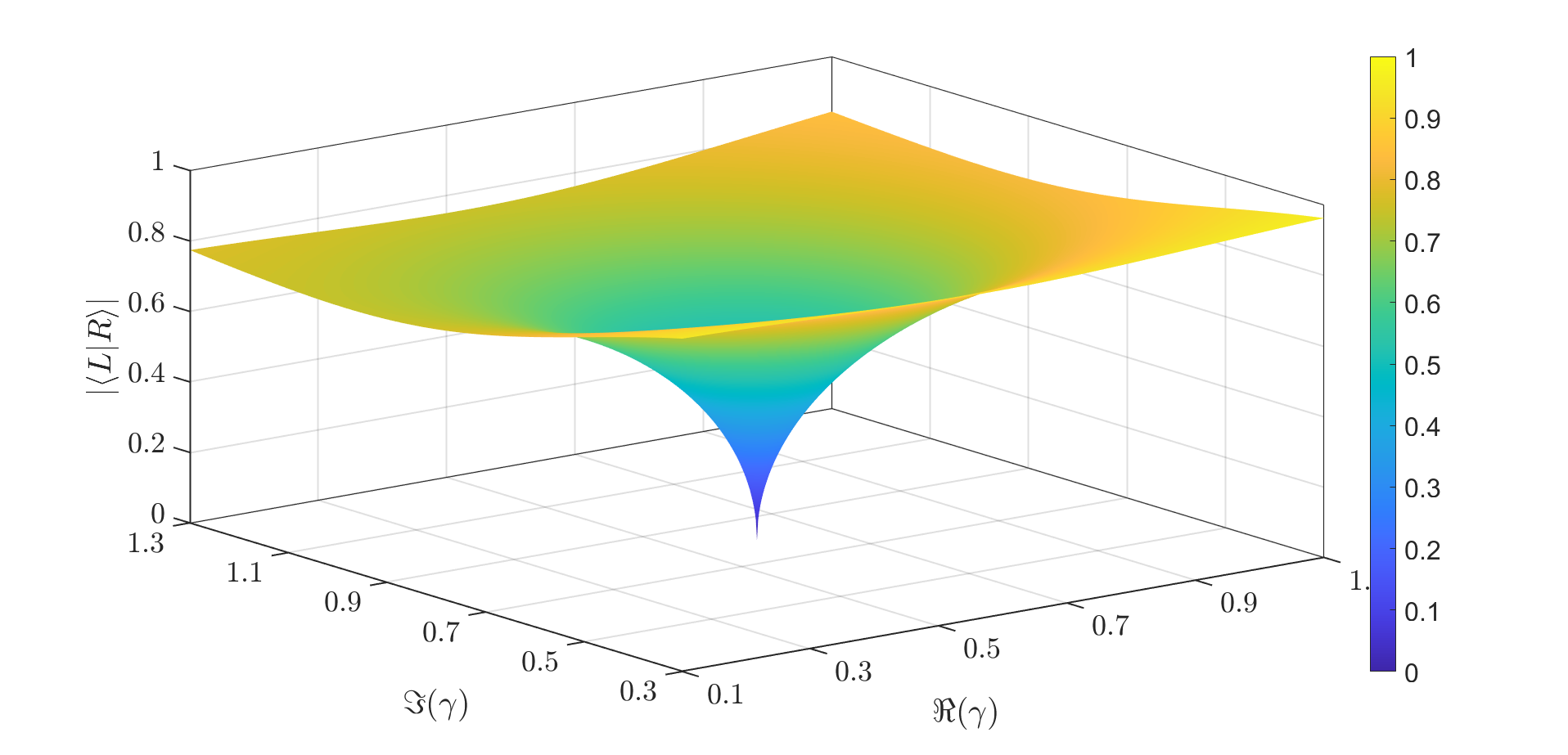}
        \caption{Plot of $|\langle L|R\rangle|$ as a function of complex $\gamma$ for $L=4$ in the $\gamma$ plane.}
        \label{figb}
\end{figure}

Recalling the definition (\ref{lambda}),
we set \(\lambda= \mathrm{i} \lambda_I\), with \(\lambda_I\in\mathbb{R}\), and obtain
\begin{equation}
    \gamma(\lambda_I)= \frac{1-\lambda_I^2}{1+\lambda_I^2} + \mathrm{i} \frac{2\lambda_I}{1+\lambda_I^2}.
\label{eq:lambda-circle}
\end{equation}
It follows that the imaginary \(\lambda\) axis is mapped to the unit circle \(|\gamma|=1\) in the complex \(\gamma\) plane. 

The exact spectral construction also gives a local measure of eigenvalue sensitivity. For a simple eigenvalue \(E\) of the complex-symmetric spin Hamiltonian \(H_\gamma\), its absolute eigenvalue condition number in the Euclidean norm is~\cite{TrefethenEmbree2005,Wiersig2023}
\begin{equation}
 \kappa_E=\frac{v^\dagger v}{|v^Tv|}=\frac{1}{|\rho|}.
 \label{eq:condition-number}
\end{equation}
For a small matrix perturbation \(\Delta H\), perturbation theory gives
\begin{equation}
 \Delta E=\frac{v^T\Delta H\,v}{v^Tv}
 +O(\|\Delta H\|_2^2),\qquad
 |\Delta E|\leq\kappa_E\|\Delta H\|_2+O(\|\Delta H\|_2^2).
 \label{eq:first-order-sensitivity}
\end{equation}
The expansion is local to a fixed simple eigenvalue and is not uniform as an EP is approached. Figures~\ref{figa} and~\ref{figb} display the inverse condition number through \(|\rho|\). Along either simple branch approaching the EP, \(|\rho|\to0\) and \(\kappa_E\to\infty\).

The \(\eta\)-pseudospectrum characterizes sensitivity to all matrix perturbations bounded in the specified norm~\cite{TrefethenEmbree2005}:
\begin{equation}
 \sigma_\eta(H_\gamma)=
 \bigcup_{\|\Delta H\|_2<\eta}\sigma(H_\gamma+\Delta H),
 \qquad \eta>0.
 \label{eq:pseudospectrum}
\end{equation}
An isolated two-dimensional Jordan block generically has square-root eigenvalue splitting under a small perturbation, rather than the linear response of a simple eigenvalue~\cite{Heiss2012,Wiersig2023}. The exact EP position and Jordan form at fixed \(\gamma\) are therefore not protected against arbitrary imperfections. The condition number bounds the worst-case first-order local response. A change of anisotropy, local disorder, and perturbations that preserve the quadratic structure need not attain the same response.

The exact solution determines the coalesced eigenvectors and Jordan chains at the EP and the square-root splitting of nearby branches. The permutation obtained by analytic continuation around an isolated EP is unchanged under sufficiently small continuous deformations for which the followed eigenvalues remain simple everywhere on the loop throughout the deformation. This parameter-space property of the spectrum is distinct from a bulk topological phase.

\section{Monodromy around exceptional points}

In the previous section, the exact solution of the non-Hermitian XY model at EPs was constructed, and the plots of the overlap $|\langle L|R\rangle|$ in Figure~\ref{figa} and Figure~\ref{figb} illustrate the self-orthogonality condition $\langle L|R\rangle=0$ at an EP. 
\label{sec:monodromy}

The relevant structure is the permutation of eigenvalues and their one-dimensional eigenspaces under analytic continuation around an EP. No bulk topological phase or bulk-boundary correspondence is implied.
Previous studies~\cite{Kawabata2019,Ding2022} showed that, upon encircling an EP in complex parameter space, eigenenergies and eigenstates are permuted.

In the present model, we observe that an analogous phenomenon arises when the anisotropy parameter $\gamma$ is varied in the complex plane. 
In this section, we provide an analytic argument for this behavior within the framework of the exactly solved XY model.

Returning to the polynomial form of the eigenvalue equations (\ref{eq:xy-poly-mode-I-boundary}) and (\ref{eq:xy-poly-mode-II-boundary}) for modes $\mathrm{I}$ and $\mathrm{II}$, we have
\begin{eqnarray}
    U_{L/2}\left(x\right)-\lambda U_{L/2-1}\left(x\right)&=&0,\label{rooteqI}\\
    U_{L/2}\left(x\right)-\lambda^{-1}U_{L/2-1}\left(x\right)&=&0.\label{rooteqII}
\end{eqnarray}

These are polynomial equations of degree ${L}/{2}$ in \(x\).
Let \(P(\varepsilon,\gamma)\) denote the corresponding boundary expression after substituting \(x=x(\varepsilon)\), in a neighborhood where its coefficients are analytic. At every EP of the finite even chain, \(P\) has a double root in \(\varepsilon\). Since \(\partial_x P_n=0\) there, differentiation at fixed \(\varepsilon\) gives \(P_\gamma=-\zeta'(\gamma_{\rm EP})U_{n-1}(x_*)\). Both choices of \(\zeta\) have nonzero derivative at the EP, and \(U_n\) and \(U_{n-1}\) have no common zero. Hence
\begin{equation}
    P=P_\varepsilon=0,\qquad P_{\varepsilon\varepsilon}\ne0,
    \qquad P_\gamma\ne0
    \quad\hbox{at }(\varepsilon_{\rm EP},\gamma_{\rm EP}).
\end{equation}
Taylor expansion then gives
\begin{equation}
    \varepsilon_\pm(\gamma)=\varepsilon_{\rm EP}
    \pm\left[-\frac{2P_\gamma}{P_{\varepsilon\varepsilon}}
    (\gamma-\gamma_{\rm EP})\right]^{1/2}
    +O(\gamma-\gamma_{\rm EP}),
\end{equation}
where the derivatives are evaluated at the EP. These conditions specify square-root splitting. Multiplicity alone does not establish this splitting or defectiveness. The latter is established by the Jordan-chain construction. For such a pair, a loop around the branch point exchanges the two quasi-energy branches. Their polynomial eigenvectors continue with the same roots. The induced permutation of many-body energies is obtained from their free-fermion sums. A sum can remain single-valued when the exchanged contributions cancel.
The coalescing quasi-energy branches and their mode eigenvectors therefore form a two-sheeted Riemann-surface structure in the complex $\gamma$-plane, as is characteristic of EP branch singularities in non-Hermitian systems~\cite{Heiss2012,Bergholtz2021,Ding2022}. 

\begin{figure}[ht]
        \centering
        \includegraphics[width=0.9\textwidth]{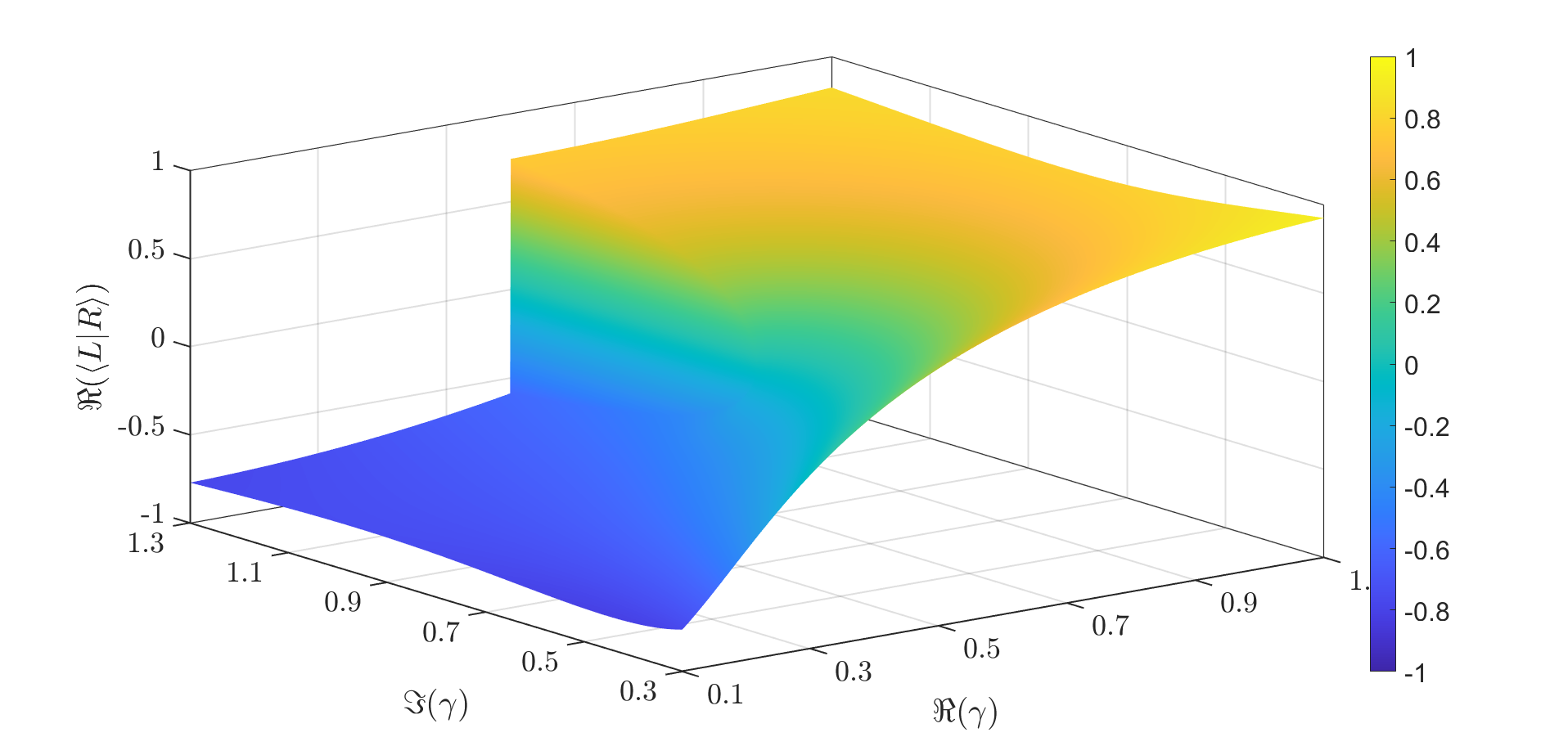}
        \includegraphics[width=0.9\textwidth]{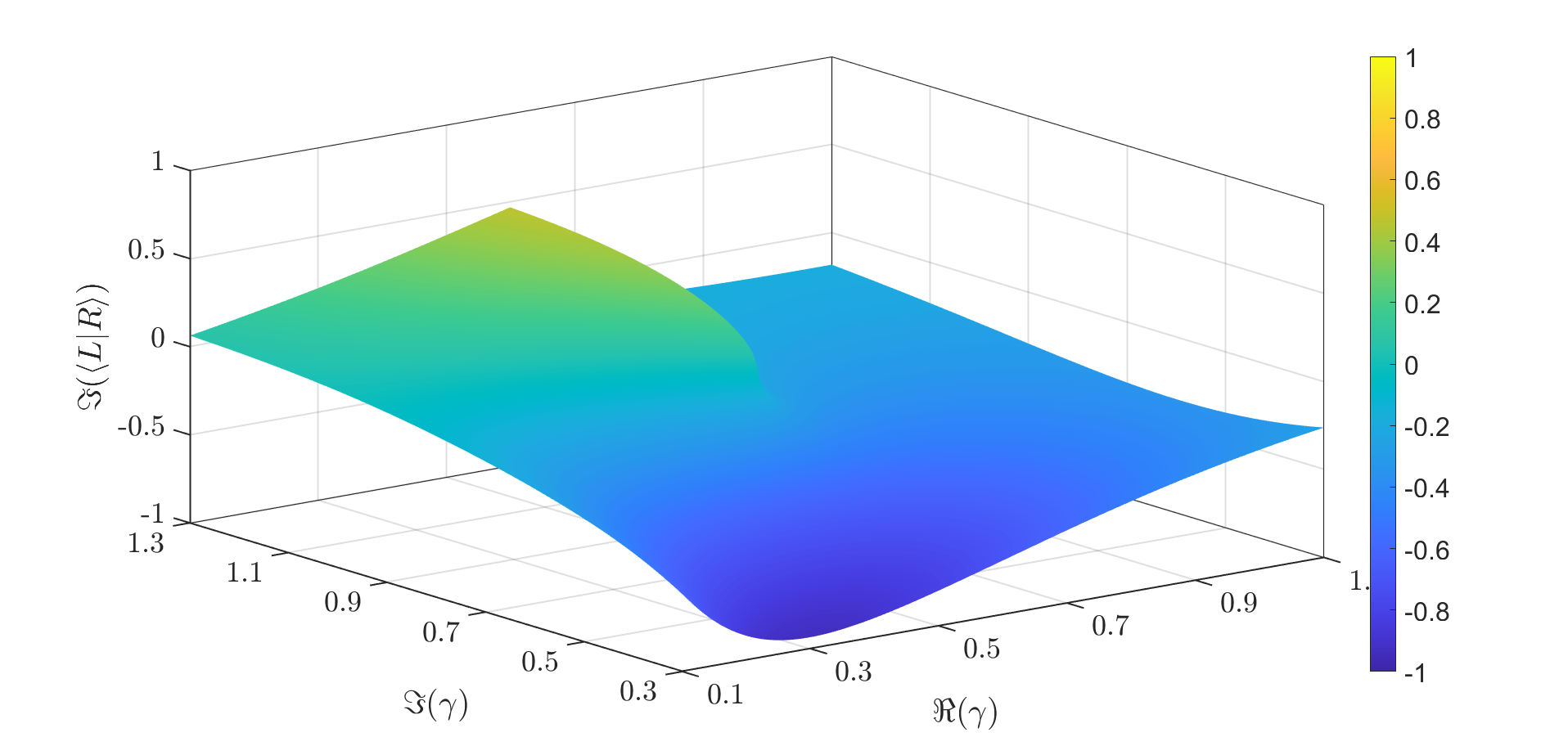}
        \caption{Plots of the real and imaginary parts of $\langle L|R\rangle$ as functions of $\gamma$ for $L=4$ in the complex $\gamma$ plane. The normalization and phase convention are specified in Eq.~\eqref{eq:overlap-normalization} and Appendix~\ref{app:L4}.}
        \label{fig1}
\end{figure}

Figure~\ref{fig1} plots the real and imaginary parts of the same normalized left-right overlap in the complex \(\gamma\)-plane.
The discontinuity of these two components across the cut gives a direct visualization of the Riemann-sheet structure of the degenerating eigenvectors.

The two sheets are connected by a chosen branch cut, which is directly related to the eigenstate permutation when encircling the EP. 
The location of the cut is conventional, but the branch point and the permutation are independent of that choice. Since the Euclidean normalization contains complex conjugation, the plotted overlap is generally not holomorphic. The algebraic Riemann surface belongs to the eigenvalue and unnormalized eigenvector expressions.

For a concrete example, consider the exact $L=4$ solution given in Appendix~\ref{app:L4} and EPs given in Appendix~\ref{EPsList}.
The two branches shown in Figure~\ref{fig:ep-eigenvector-exchange} can be chosen as the $\eta=+1$ branch of $E^{(5)}_{\eta,s}$, namely
\begin{eqnarray}
    D_{\pm}(\gamma)&=&5\gamma^2\pm6\gamma+5, \nonumber\\
    E_{s}^{\rm ex}(\gamma)&=&-\frac{\sqrt{D_+(\gamma)}+s\sqrt{D_-(\gamma)}}{4},\qquad
    v_{s}^{\rm ex}=v^{(5)}_{+,s},\qquad s=\pm1 .
\label{eq:L4-monodromy-pair}
\end{eqnarray}
At $\gamma_{\rm EP}^{(+)}=\frac{3}{5}+\frac{4}{5}\mathrm{i}$, $D_-(\gamma_{\rm EP}^{(+)})=0$ while $D_+(\gamma_{\rm EP}^{(+)})\neq0$. The other zero of $D_-$ is $\gamma_{\rm EP}^{(-)}=\frac{3}{5}-\frac{4}{5}\mathrm{i}$. 

The simple zero \(D_-'(\gamma_{\rm EP}^{(+)})=8\mathrm{i}\neq0\) makes \(\sqrt{D_-}\) change sign after a circuit of this EP, while \(\sqrt{D_+}\) is locally single-valued. Thus the two energies and their eigenvectors are exchanged.
To see the chosen branch cut in Figure~\ref{fig:ep-eigenvector-exchange}, take two points on the two sides of the cut,
\begin{eqnarray}
    \gamma_{\pm}=\gamma_{\rm EP}^{(+)}+(\pm r+\mathrm{i}R),\qquad 0<r\ll1,\quad R>0.
\end{eqnarray}
Using $D_-(\gamma)=5(\gamma-\gamma_{\rm EP}^{(+)})(\gamma-\gamma_{\rm EP}^{(-)})$, one obtains, to leading order in $r$,
\begin{eqnarray}
    D_-(\gamma_{\pm})&=&-R(8+5R)\pm \mathrm{i}(8+10R)r+O(r^2) .
\end{eqnarray}
Thus the infinitesimal displacement $\pm r$ selects the two boundary values of the square root. 
On the branch-cut ray shown in Figure~\ref{fig:ep-eigenvector-exchange}, where the real limiting value of $D_-$ lies on the chosen cut of the square root, this gives a finite phase jump. The arguments of $D_-(\gamma_+)$ and $D_-(\gamma_-)$ differ by $2\pi$ in the limit $r\rightarrow0^+$, and hence the two boundary values of $\sqrt{D_-(\gamma)}$ differ by a phase $\pi$, i.e., by a sign. 
Consequently the two boundary values of the exact energy are exchanged,
\begin{eqnarray}
    \lim_{r\rightarrow0^+}E_{s}^{\rm ex}(\gamma_+)&=&\lim_{r\rightarrow0^+}E_{-s}^{\rm ex}(\gamma_-) .
\end{eqnarray}
The same discontinuity is inherited by the eigenvectors, because the explicit components of $v^{(5)}_{+,s}$ in Appendix~\ref{app:L4} are rational functions of $\gamma$ and $E_{s}^{\rm ex}$, and therefore $E_s^{\rm ex}$ and $v_s^{\rm ex}$ belong to the same algebraic extension field $\mathbb{Q}(\gamma,\sqrt{D_+(\gamma)},\sqrt{D_-(\gamma)})$. 
\label{par:field-extension}

In these explicit expressions, the two eigenvectors coalesce when \(D_-=0\). The Jordan construction in Section~\ref{sec:EP}, and the direct rank check in Appendix~\ref{app:L4}, establish an EP rather than a diabolic point and membership in a common field extension alone would not establish defectiveness.
This exact $L=4$ solution therefore gives the chosen branch cut in Figure~\ref{fig:ep-eigenvector-exchange}. Crossing the ray jumps from one boundary value of the algebraic square root to the other and exchanges the corresponding eigenstate.

In the present formulation, this branch structure is already visible before solving for \(k\). It is the monodromy of the roots of the quasi-energy boundary polynomial, and the polynomial eigenvectors inherit the same monodromy through \(x(\varepsilon)\).

The same relation holds for \(\rho_s\) in the fixed phase convention. Crossing the cut changes the branch label used to describe analytic continuation. It is not a physical discontinuity of the Hamiltonian.
\begin{figure}[ht]
        \centering
        \includegraphics[width=0.9\textwidth]{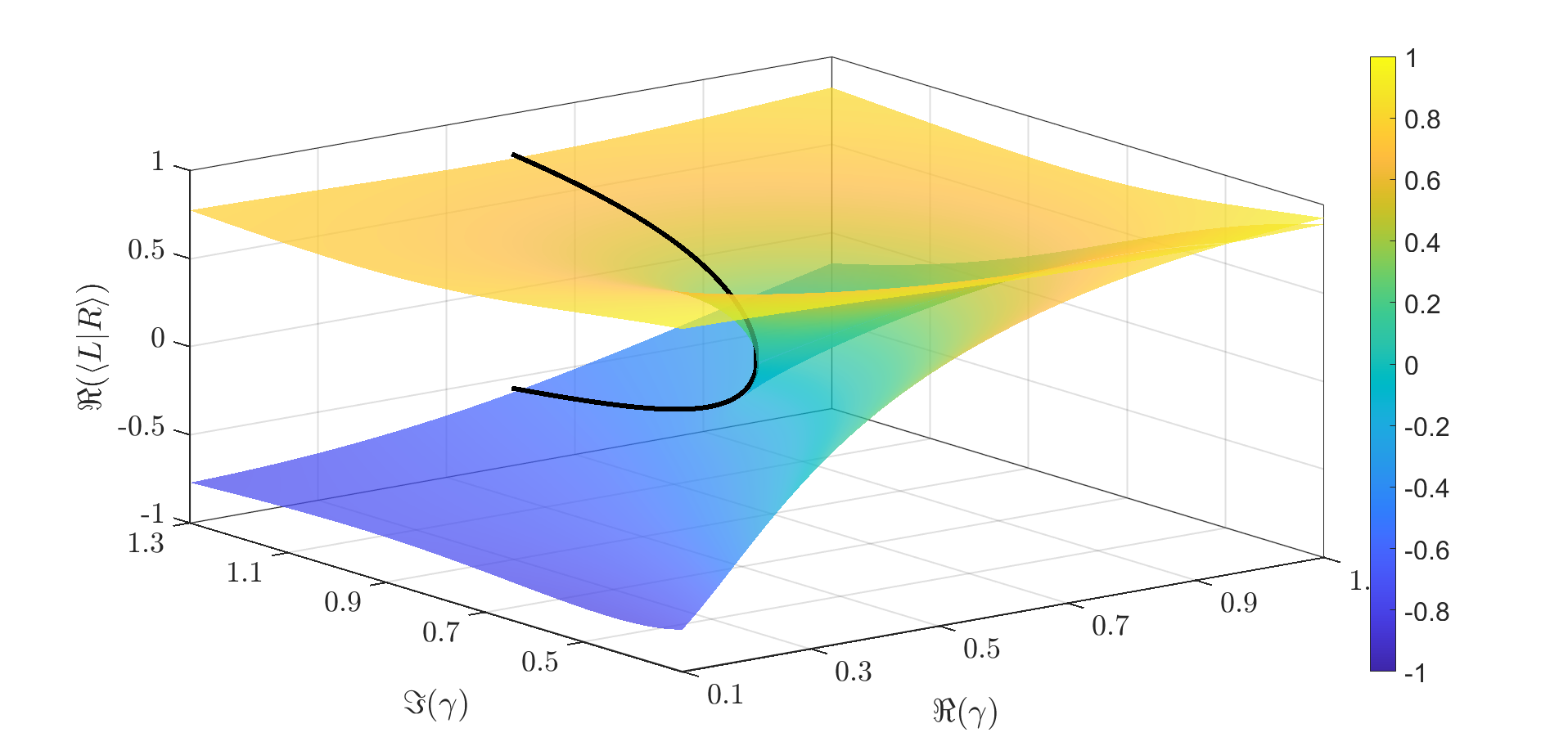}
        \includegraphics[width=0.9\textwidth]{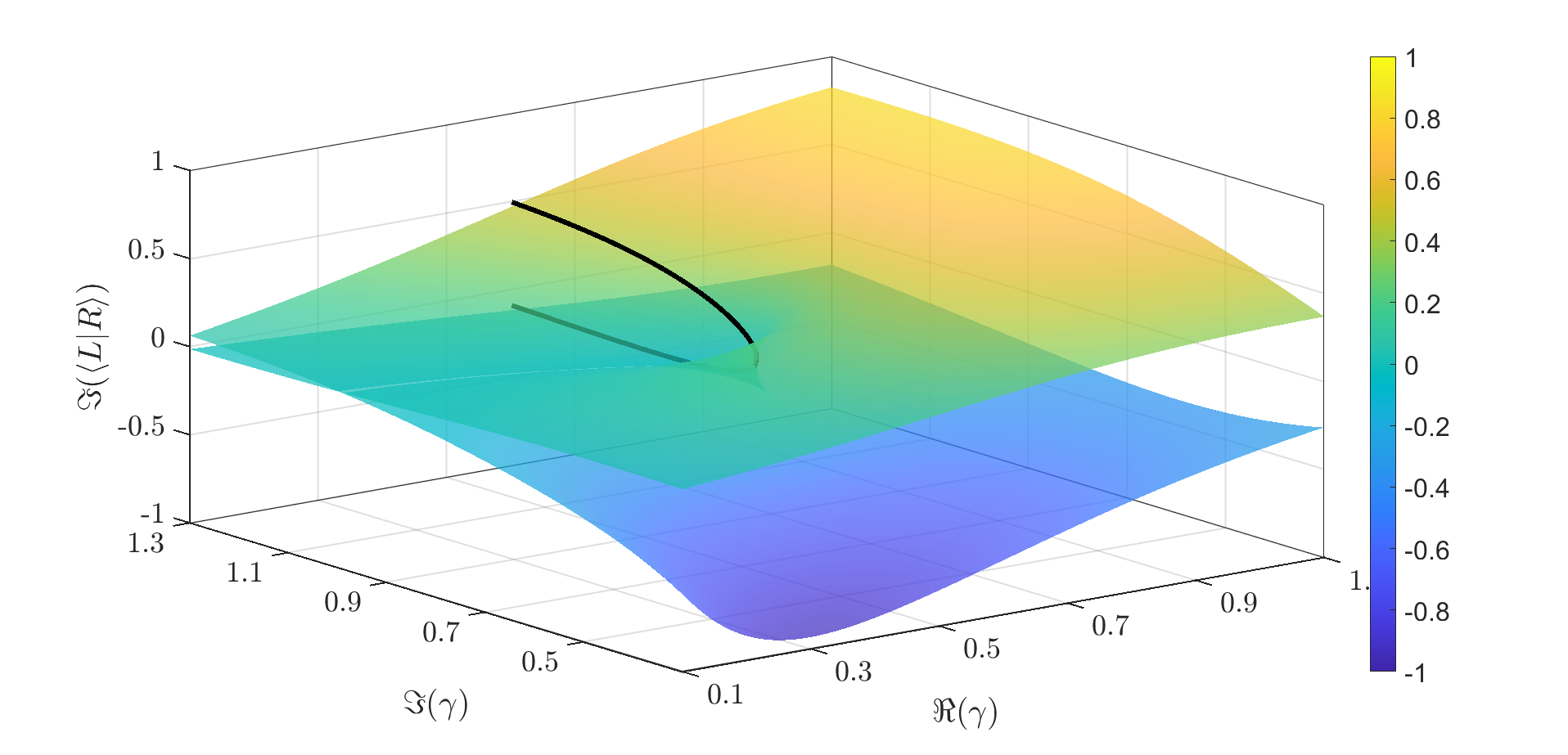}
        \caption{Real and imaginary parts of the normalized overlap $\langle L|R\rangle$ for the pair of degenerating eigenvectors that exchange when the EP is encircled in the complex $\gamma$ plane. The black line marks the chosen branch cut. Crossing it changes the branch label used for analytic continuation.}
        \label{fig:ep-eigenvector-exchange}
\end{figure}

To make this exchange more explicit at the level of the two degenerating branches, Figure~\ref{fig:ep-eigenvector-exchange} shows the real and imaginary parts of the stitched two-sheet surface of $\langle L|R\rangle$ for the eigenvector pair that is permuted by encircling the EP.
The seam between the two sheets visualizes the branch-cut structure. After one loop around the EP, analytic continuation carries the eigenvector onto the other sheet.
These observations show that, upon encircling an EP, both eigenenergies and eigenstates undergo a non-trivial permutation for the same algebraic reason. 

This continuation concerns the instantaneous spectral problem. Actual evolution with a time-dependent \(\gamma(t)\) requires solving the evolution equation and need not follow a single eigenvalue branch. The chiral state transfer associated with dynamical encircling of Liouvillian EPs is a distinct open-system evolution problem~\cite{SunYi2024LEP}.

\section{Conclusion}
\label{sec:conclusion}

We have analyzed the non-Hermitian XY spin chain with open boundary conditions and shown that it remains exactly solvable within the free-fermion framework when the anisotropy parameter is extended into the complex plane.
By formulating the problem in terms of the quasi-Hamiltonian matrix, we demonstrated that the structure of the quasi-energy spectrum coincides with that of the Hermitian model, and we constructed an explicit biorthogonal fermionic basis away from EPs.
A central technical result is the Chebyshev-polynomial form of the open-boundary eigenvectors.
In this form the quasi-energy \(\varepsilon\), rather than the quasi-momentum \(k\), is the natural variable, and the familiar trigonometric solution is recovered after setting \(x(\varepsilon)=\cos 2k\).
A primary motivation for this work is that at EPs, where the quasi-Hamiltonian becomes defective, the standard diagonalization procedure breaks down.
We showed that the defective eigensystem can nevertheless be reconstructed explicitly.
The polynomial representation makes this step algebraic. Since EPs are repeated roots of the quasi-energy boundary polynomial, the generalized eigenvectors arise naturally from derivatives of the polynomial eigenvectors with respect to \(\varepsilon\), retaining the boundary residual before imposing the repeated-root condition.
For finite even $L$, the Chebyshev structure proves that every exceptional point is an EP2, with a single defective mode family and all remaining quasiparticle modes diagonalizable. The many-body lift gives \(2^{L-2}\) blocks of size two and \(2^{L-1}\) blocks of size one, or equivalently \(3\times2^{L-2}\) independent eigenvectors and \(2^{L-2}\) generalized eigenvectors.

We further investigated the monodromy around EPs by analyzing the analytic structure of the quasi-energies and eigenvectors as functions of the complex anisotropy parameter.
Using a Chebyshev-polynomial formulation of the quantization condition, we identified branch-point singularities in parameter space and showed that encircling an isolated second-order EP leads to a permutation of eigenenergies and eigenstates.
This provides a real-space illustration of familiar EP monodromy and is consistent with the branch-cut structure observed in the normalized left-right overlap in the fixed phase convention used here.
Our results establish the non-Hermitian XY chain with open boundaries as an analytically controlled quadratic many-body system in which EP physics, biorthogonal structures, and spectral monodromy can be studied beyond translationally invariant settings.
More broadly, the advantage of the quasi-energy polynomial method is that the spectral equation, the EP multiple-root condition, and the eigenvector continuation are encoded using the same quasi-energy variable.
Extensions to interacting or parafermionic systems require an independent construction. The overlap modulus measures the inverse condition number along a simple eigenvalue branch, identifying enhanced local sensitivity near the EP.

\label{par:implementation}
Physical spin expectation values in a conditional experiment require a normalized prepared state and a specified measurement protocol. In conditional no-jump realizations, spatially extensive loss can make the probability of retaining a trajectory decrease exponentially with size and time~\cite{Ashida2020}. Variational quantum simulations of other non-Hermitian Hamiltonians provide examples of access to selected dynamics without direct trajectory postselection, including the trapped-ion study of Zhang \textit{et al.}~\cite{Zhang2025TrappedIon}. A realization of the present model requires specifying its state preparation, measurement procedure, and, for an open-system implementation, its system--environment coupling.

\section*{Acknowledgements}
This work was supported by
Australian Research Council Grant DP240100838.

\appendix
\numberwithin{equation}{section}

\section{Anticommutation relations of \(L^*\) and \(R\) operators and the Jordan basis}
\label{app:CAR}

For the system at non-EPs, the matrix \(V\) is orthogonal, so \(V^{-1}=V^{T}\). 
Consequently, the coefficients appearing in the inverse transformation have the same components as those in \(P\), i.e.,  the symbols \(\phi^{-1}\) and \(\psi^{-1}\) may be identified componentwise with \(\phi\) and \(\psi\) (up to the transpose implied by the matrix notation).
Using this fact, together with the canonical anticommutation relations of \(c_\mu\) and \(c_\mu^\dagger\), we can verify the anticommutation relations of the biorthogonal operators explicitly. 
Several representative calculations are presented below. 
In each case, the procedure is the same. First expand the operator products, then use \(\{c_\mu,c_\nu\}=0\), \(\{c_\mu^\dagger,c_\nu^\dagger\}=0\), and \(\{c_\mu,c_\nu^\dagger\}=\delta_{\mu\nu}\), and finally apply the orthogonality relations of \(\phi\) and \(\psi\). For example, 
\begin{eqnarray}
    \{L^*_i, R_j\}&=&\frac{1}{2}\left\{\phi_{\mathrm{I}i}^{\mu}(c_\mu+c_\mu^\dagger)-\psi_{\mathrm{I}i}^{\mu}(c_\mu-c_\mu^\dagger),\phi_{\mathrm{I}j}^{-1\nu}(c_\nu+c_\nu^\dagger)+\psi_{\mathrm{I}j}^{-1\nu}(c_\nu-c_\nu^\dagger)\right\}\\
    &=&\phi_{\mathrm{I}i}^{\mu}\phi_{\mathrm{I}j}^{-1\nu}\delta_{\mu\nu}+\psi_{\mathrm{I}i}^{\mu}\psi_{\mathrm{I}j}^{-1\nu}\delta_{\mu\nu}\nonumber\\
    &=&\delta_{ij} \,.\nonumber
\end{eqnarray}
Here the mixed terms cancel after using the canonical anticommutation relations, and the remaining contraction terms reduce to \(\delta_{ij}\) by the orthogonality condition in Eq.~(\ref{orth}). Also 
\begin{eqnarray}
    \{L^*_i,L_j\}&=&\frac{1}{2}\left\{\phi_{\mathrm{I}i}^{\mu}(c_\mu+c_\mu^\dagger)-\psi_{\mathrm{I}i}^{\mu}(c_\mu-c_\mu^\dagger),\phi_{\mathrm{II}j}^{\nu}(c_\nu+c_\nu^\dagger)-\psi_{\mathrm{II}j}^{\nu}(c_\nu-c_\nu^\dagger)\right\}\\
    &=&\phi_{\mathrm{I}i}^{\mu}\phi_{\mathrm{II}j}^{\nu}\delta_{\mu\nu}-\psi_{\mathrm{I}i}^{\mu}\psi_{\mathrm{II}j}^{\nu}\delta_{\mu\nu}\nonumber\\
    &=&0 \,.\nonumber
\end{eqnarray}
This vanishing follows from the explicit parity structure of the type-\(\mathrm{I}\) and type-\(\mathrm{II}\) mode functions. The nonzero components of \(\phi_{\mathrm{I}}\) and \(\phi_{\mathrm{II}}\), and also those of \(\psi_{\mathrm{I}}\) and \(\psi_{\mathrm{II}}\), lie on opposite sublattices.
As another example, 
\begin{eqnarray}
    \{L^*_i,R^*_j\}&=&\frac{1}{2}\left\{\phi_{\mathrm{I}i}^{\mu}(c_\mu+c_\mu^\dagger)-\psi_{\mathrm{I}i}^{\mu}(c_\mu-c_\mu^\dagger),\phi_{\mathrm{II}j}^{-1\nu}(c_\nu+c_\nu^\dagger)+\psi_{\mathrm{II}j}^{-1\nu}(c_\nu-c_\nu^\dagger)\right\}\\
    &=&\phi_{\mathrm{I}i}^{\mu}\phi_{\mathrm{II}j}^{-1\nu}\delta_{\mu\nu}+\psi_{\mathrm{I}i}^{\mu}\psi_{\mathrm{II}j}^{-1\nu}\delta_{\mu\nu}\nonumber\\
    &=&0 \,.\nonumber
\end{eqnarray}
The same cancellation mechanism applies here. After reducing to \(\delta_{\mu\nu}\), the coefficient combination vanishes by the orthogonality relations.
Finally
\begin{eqnarray}
    R_iR_i&=&\frac{1}{2}\left[\phi_{\mathrm{I}i}^{-1\mu}(c_\mu+c_\mu^\dagger)+\psi_{\mathrm{I}i}^{-1\mu}(c_\mu-c_\mu^\dagger)\right]\left[\phi_{\mathrm{I}i}^{-1\nu}(c_\nu+c_\nu^\dagger)+\psi_{\mathrm{I}i}^{-1\nu}(c_\nu-c_\nu^\dagger)\right]\\
    &=&\frac{1}{2}\bigg[\phi_{\mathrm{I}i}^{-1\mu}\phi_{\mathrm{I}i}^{-1\nu}(c_\mu c_\nu+ c_\mu^\dagger c_\nu+ c_\mu c_\nu^\dagger+ c_\mu^\dagger c_\nu^\dagger)+\psi_{\mathrm{I}i}^{-1\mu}\phi_{\mathrm{I}i}^{-1\nu}(c_\mu c_\nu- c_\mu^\dagger c_\nu+ c_\mu c_\nu^\dagger- c_\mu^\dagger c_\nu^\dagger)+\nonumber\\
    &&\phi_{\mathrm{I}i}^{-1\mu}\psi_{\mathrm{I}i}^{-1\nu}(c_\mu c_\nu+ c_\mu^\dagger c_\nu- c_\mu c_\nu^\dagger- c_\mu^\dagger c_\nu^\dagger)+\psi_{\mathrm{I}i}^{-1\mu}\psi_{\mathrm{I}i}^{-1\nu}(c_\mu c_\nu- c_\mu^\dagger c_\nu- c_\mu c_\nu^\dagger+ c_\mu^\dagger c_\nu^\dagger)\bigg]\nonumber\\
    &=&\frac{1}{2}\bigg[\left(\phi_{\mathrm{I}i}^{-1\mu}\phi_{\mathrm{I}i}^{-1\nu}+(\psi_{\mathrm{I}i}^{-1\mu}\phi_{\mathrm{I}i}^{-1\nu}+\phi_{\mathrm{I}i}^{-1\mu}\psi_{\mathrm{I}i}^{-1\nu})+\psi_{\mathrm{I}i}^{-1\mu}\psi_{\mathrm{I}i}^{-1\nu}\right)c_\mu c_\nu+\nonumber\\
    &&\left(\phi_{\mathrm{I}i}^{-1\mu}\phi_{\mathrm{I}i}^{-1\nu}-(\psi_{\mathrm{I}i}^{-1\mu}\phi_{\mathrm{I}i}^{-1\nu}+\phi_{\mathrm{I}i}^{-1\mu}\psi_{\mathrm{I}i}^{-1\nu})+\psi_{\mathrm{I}i}^{-1\mu}\psi_{\mathrm{I}i}^{-1\nu}\right)c_\mu^\dagger c_\nu^\dagger+\nonumber\\
    &&\left(\phi_{\mathrm{I}i}^{-1\mu}\phi_{\mathrm{I}i}^{-1\nu}-\psi_{\mathrm{I}i}^{-1\mu}\phi_{\mathrm{I}i}^{-1\nu}+\phi_{\mathrm{I}i}^{-1\mu}\psi_{\mathrm{I}i}^{-1\nu}-\psi_{\mathrm{I}i}^{-1\mu}\psi_{\mathrm{I}i}^{-1\nu}\right)c_\mu^\dagger c_\nu+\nonumber\\
    &&\left(\phi_{\mathrm{I}i}^{-1\nu}\phi_{\mathrm{I}i}^{-1\mu}-\phi_{\mathrm{I}i}^{-1\nu}\psi_{\mathrm{I}i}^{-1\mu}+\psi_{\mathrm{I}i}^{-1\nu}\phi_{\mathrm{I}i}^{-1\mu}-\psi_{\mathrm{I}i}^{-1\nu}\psi_{\mathrm{I}i}^{-1\mu}\right)c_\nu c_\mu^\dagger\bigg]\nonumber\\
    &=&\frac{1}{4}\bigg[\left(\phi_{\mathrm{I}i}^{-1\mu}\phi_{\mathrm{I}i}^{-1\nu}+(\psi_{\mathrm{I}i}^{-1\mu}\phi_{\mathrm{I}i}^{-1\nu}+\phi_{\mathrm{I}i}^{-1\mu}\psi_{\mathrm{I}i}^{-1\nu})+\psi_{\mathrm{I}i}^{-1\mu}\psi_{\mathrm{I}i}^{-1\nu}\right)\{c_\mu,c_\nu\}+\nonumber\\
    &&\left(\phi_{\mathrm{I}i}^{-1\mu}\phi_{\mathrm{I}i}^{-1\nu}-(\psi_{\mathrm{I}i}^{-1\mu}\phi_{\mathrm{I}i}^{-1\nu}+\phi_{\mathrm{I}i}^{-1\mu}\psi_{\mathrm{I}i}^{-1\nu})+\psi_{\mathrm{I}i}^{-1\mu}\psi_{\mathrm{I}i}^{-1\nu}\right)\{c_\mu^\dagger,c_\nu^\dagger\}+\nonumber\\
    &&2\left(\phi_{\mathrm{I}i}^{-1\mu}\phi_{\mathrm{I}i}^{-1\nu}-\psi_{\mathrm{I}i}^{-1\mu}\phi_{\mathrm{I}i}^{-1\nu}+\phi_{\mathrm{I}i}^{-1\mu}\psi_{\mathrm{I}i}^{-1\nu}-\psi_{\mathrm{I}i}^{-1\mu}\psi_{\mathrm{I}i}^{-1\nu}\right)\{c_\mu^\dagger,c_\nu\}\bigg]\nonumber\\
    &=&\frac{1}{2}\left(\phi_{\mathrm{I}i}^{-1\mu}\phi_{\mathrm{I}i}^{-1\nu}-\psi_{\mathrm{I}i}^{-1\mu}\phi_{\mathrm{I}i}^{-1\nu}+\phi_{\mathrm{I}i}^{-1\mu}\psi_{\mathrm{I}i}^{-1\nu}-\psi_{\mathrm{I}i}^{-1\mu}\psi_{\mathrm{I}i}^{-1\nu}\right)\delta_{\mu\nu}\nonumber\\
    &=&0 \,.\nonumber
\end{eqnarray}
In this derivation, the terms proportional to \(\{c_\mu,c_\nu\}\) and \(\{c_\mu^\dagger,c_\nu^\dagger\}\) vanish identically, and only the \(\{c_\mu^\dagger,c_\nu\}\) contraction survives. 
The remaining coefficient then cancels, giving \(R_iR_i=0\), as required of fermionic operators.
We next consider the anticommutator between the same kind of operators. For example,
\begin{eqnarray}
    \{L_i,L_j\}&=&\frac{1}{2}\left\{\phi_{\mathrm{II}i}^{\mu}(c_\mu+c_\mu^\dagger)-\psi_{\mathrm{II}i}^{\mu}(c_\mu-c_\mu^\dagger),\phi_{\mathrm{II}j}^{\nu}(c_\nu+c_\nu^\dagger)-\psi_{\mathrm{II}j}^{\nu}(c_\nu-c_\nu^\dagger)\right\} \nonumber\\
    &=&\phi_{\mathrm{II}i}^{\mu}\phi_{\mathrm{II}j}^{\nu}\delta_{\mu\nu}-\psi_{\mathrm{II}i}^{\mu}\psi_{\mathrm{II}j}^{\nu}\delta_{\mu\nu} \nonumber\\
    &=&\phi_{\mathrm{II}i}^{T}\phi_{\mathrm{II}j}-\psi_{\mathrm{II}i}^{T}\psi_{\mathrm{II}j} .
    \label{LLapp}
\end{eqnarray}
The type-\(\mathrm{II}\) mode functions satisfy
\begin{eqnarray}
    (A-B)\psi_{\mathrm{II}j}=\varepsilon_{\mathrm{II}j}\phi_{\mathrm{II}j},\qquad (A+B)\phi_{\mathrm{II}j}=\varepsilon_{\mathrm{II}j}\psi_{\mathrm{II}j}.
\end{eqnarray}
Since \(A^T=A\) and \(B^T=-B\), it follows that
\begin{eqnarray}
    \varepsilon_{j}^{\mathrm{II}}\left(\phi_{\mathrm{II}i}^{T}\phi_{\mathrm{II}j}-\psi_{\mathrm{II}i}^{T}\psi_{\mathrm{II}j}\right)&=&\phi_{\mathrm{II}i}^{T}(A-B)\psi_{\mathrm{II}j}-\psi_{\mathrm{II}i}^{T}(A+B)\phi_{\mathrm{II}j} \nonumber\\
    \varepsilon_{j}^{\mathrm{II}}\left(\phi_{\mathrm{II}i}^{T}\phi_{\mathrm{II}j}-\psi_{\mathrm{II}i}^{T}\psi_{\mathrm{II}j}\right)&=&\varepsilon_{i}^{\mathrm{II}}\psi_{\mathrm{II}i}^{T}\psi_{\mathrm{II}j}-\varepsilon_{i}^{\mathrm{II}}\phi_{\mathrm{II}i}^{T}\phi_{\mathrm{II}j} \nonumber\\
    \varepsilon_{j}^{\mathrm{II}}\left(\phi_{\mathrm{II}i}^{T}\phi_{\mathrm{II}j}-\psi_{\mathrm{II}i}^{T}\psi_{\mathrm{II}j}\right)&=&-\varepsilon_{i}^{\mathrm{II}}\left(\phi_{\mathrm{II}i}^{T}\phi_{\mathrm{II}j}-\psi_{\mathrm{II}i}^{T}\psi_{\mathrm{II}j}\right) \nonumber\\
    \left(\varepsilon_{i}^{\mathrm{II}}+\varepsilon_{j}^{\mathrm{II}}\right)\left(\phi_{\mathrm{II}i}^{T}\phi_{\mathrm{II}j}-\psi_{\mathrm{II}i}^{T}\psi_{\mathrm{II}j}\right)&=&0 .
\end{eqnarray}
Thus \(\{L_i,L_j\}=0\) unless \(i\) and \(j\) label a pair of opposite quasi-energies.

With the pairing $v_{-k}=Tv_{+k}$ used in Eq.~(\ref{eq:canonical-modes}), the opposite-energy partners are \(2m-1\) and \(2m\). 
Using Eq.~(\ref{orth}), one then finds
\begin{eqnarray}
    \{L_{2m-1},L_{2n-1}\}&=&0,\nonumber\\
    \{L_{2m},L_{2n}\}&=&0,\nonumber\\
    \{L_{2m-1},L_{2n}\}&=&\left(\phi_{\mathrm{II},2m-1}^{T}\phi_{\mathrm{II},2n}-\psi_{\mathrm{II},2m-1}^{T}\psi_{\mathrm{II},2n}\right)I \nonumber\\
    &=&+\left(\phi_{\mathrm{II},2m-1}^{T}\phi_{\mathrm{II},2n-1}+\psi_{\mathrm{II},2m-1}^{T}\psi_{\mathrm{II},2n-1}\right)I \nonumber\\
    &=&+\delta_{mn}I .
\end{eqnarray}
The remaining relations, namely \(\{L_i, R^*_j\}=\delta_{ij}\), \(\{R_i,R^*_j\}=0\), \(\{L_i,R_j\}=0\), \(R_i^* R_i^*=0\) and the canonical anticommutation relations of \(\{L^*_i,L^*_j\}\), \(\{R_i,R_j\}\), \(\{R^*_i,R^*_j\}\), are verified in exactly the same way. Expand the operator products, reduce them using the canonical anticommutation relations of \(c_\mu\) and \(c_\mu^\dagger\), and then apply the orthogonality identities for the coefficient vectors.

For the Nambu column $C$ and $T=\left(\begin{smallmatrix}0&I\\I&0\end{smallmatrix}\right)$, the original canonical relations imply
\begin{equation}
 \{C_\mu,C_\nu\}=T_{\mu\nu},\qquad
 \{C_\mu,(C^\dagger)_\nu\}=\delta_{\mu\nu},\qquad C^\dagger=C^TT.
\end{equation}
For any invertible transformation $V$, including a full Jordan transformation,
\begin{equation}
 \begin{aligned}
 \mathcal R&=V^{-1}C,\qquad \mathcal L=C^\dagger V,\\
 \{\mathcal R_i,\mathcal L_j\}&=\delta_{ij},\\
 \{\mathcal R_i,\mathcal R_j\}&=(V^{-1}TV^{-T})_{ij},\qquad
 \{\mathcal L_i,\mathcal L_j\}=(V^TTV)_{ij}.
 \end{aligned}
 \label{eq:full-CAR-transform}
\end{equation}
The inverse superscripts on $\phi$ and $\psi$ denote the corresponding blocks of $V^{-1}S$, as defined in the main text. Equation~(\ref{eq:full-CAR-transform}) distinguishes the mixed dual relation, which follows from invertibility alone, from the remaining relations, which depend on the pairing of the basis vectors.

For the independent canonical operators in the main text, choose opposite-energy partners \(v_{-k}=Tv_{+k}\). With \(V^TV=I\), this gives \(\{f_i,\bar f_j\}=\delta_{ij}\), while the anticommutators within either set vanish. The signed mode operators themselves include both partners and must not all be treated as mutually anticommuting annihilation operators.
At a nonzero length-two EP, symmetry of $M$ implies
\begin{equation}
 u_0^Tu_0=u_0^T(M-\varepsilon_{EP}I)u_1
 =[(M-\varepsilon_{EP}I)u_0]^Tu_1=0.
\end{equation}
For a single length-two block, the bilinear pairing on its generalized eigenspace is nondegenerate, so $d=u_0^Tu_1\neq0$. A common multiplication of $u_0,u_1$ by $d^{-1/2}$ followed by a shift of $u_1$ makes their Gram matrix $\left(\begin{smallmatrix}0&1\\1&0\end{smallmatrix}\right)$, as in Eq.~(\ref{eq:Jordan-bilinear-normalization}). Since $T$ maps this generalized eigenspace at $+\varepsilon_{EP}$ to the one at $-\varepsilon_{EP}$, and generalized eigenspaces of a symmetric matrix for distinct eigenvalues are bilinearly orthogonal,
\begin{equation}
 u_i^TTu_j=0,\qquad i,j=0,1.
\end{equation}
Substitution into Eq.~(\ref{eq:Jordan-canonical-modes}) gives $\{f_i,\bar f_j\}=\delta_{ij}$ and zero anticommutators within either set. The unaffected generalized eigenspaces are orthogonal to this pair, so their canonical modes anticommute with these operators as well.

Finally, Eq.~(\ref{eq:nambu-normalization}) gives $[H_\gamma,C]=MC$. Therefore
\begin{equation}
 [H_\gamma,\bar f_1]=\varepsilon_{EP}\bar f_1,\qquad
 [H_\gamma,\bar f_2]=\varepsilon_{EP}\bar f_2+\bar f_1.
\end{equation}
The canonical relations then identify the defective contribution as
$\varepsilon_{EP}(\bar f_1f_1+\bar f_2f_2-1)+\bar f_1f_2$.
Its constant is fixed by $\operatorname{Tr}H_\gamma=0$, consistently with the $-1/2$ term for each ordinary mode. This proves Eq.~(\ref{Ham_ep}) with the physical Hamiltonian normalization and supplies a complete basis after adjoining the generalized states.

\section{Analytic \(L=4\) spectrum and eigenvectors}
\label{app:L4}

In this Appendix we give the exact energy eigenvalues and eigenvectors for the \(L=4\) open chain. This example makes explicit how the polynomial construction in the main text reproduces the full energy spectrum and the corresponding eigenvectors. 
The vectors are expressed in the ordered \(\sigma^z\) product basis \((|0000\rangle,|0001\rangle,\ldots,|1111\rangle)\), where \(\sigma_j^z|s_j\rangle=(-1)^{s_j}|s_j\rangle\).
For \(L=4\), the two boundary polynomials reduce to quadratic equations. Using \(U_1(x)=2x\), \(U_2(x)=4x^2-1\), Eqs.~(\ref{eq:xy-poly-mode-I-boundary}) and~(\ref{eq:xy-poly-mode-II-boundary}) become 
\begin{eqnarray}
    4x^2-1-2\lambda x &=& 0,\\
    4x^2-1-2\lambda^{-1}x &=& 0.
\end{eqnarray}
Substituting \(x=x(\varepsilon)\) gives the quasi-energy roots in radical form. It is convenient to introduce
\begin{eqnarray}
    D_{\pm}(\gamma)&=&5\gamma^2\pm6\gamma+5.
\end{eqnarray}
The quasi-energies may then be written as
\begin{eqnarray}
    \varepsilon^{\mathrm{I}}_{\eta,s}=\frac{\eta(1-\gamma)+s\sqrt{D_+(\gamma)}}{4},\quad \varepsilon^{\mathrm{II}}_{\eta,s}=\frac{\eta(1+\gamma)+s\sqrt{D_-(\gamma)}}{4},\qquad \eta,s=\pm1 .
\end{eqnarray}
Combining these quasi-energies according to the free-fermion construction gives the full \(2^4\)-dimensional many-body spectrum. 
For the normalization of \(H_\gamma\) used in the main text, the sixteen eigenvalues are
\begin{eqnarray}
    \{E\}&=&\left\{\pm\frac{1}{2},\pm\frac{\gamma}{2},\frac{1-\gamma\pm\sqrt{D_-(\gamma)}}{4},\frac{\gamma-1\pm\sqrt{D_-(\gamma)}}{4},-\frac{1+\gamma}{4}\pm\frac{\sqrt{D_+(\gamma)}}{4},\frac{1+\gamma}{4}\pm\frac{\sqrt{D_+(\gamma)}}{4},\right.\nonumber\\
    &&\left.\pm\frac{\sqrt{D_+(\gamma)}+\sqrt{D_-(\gamma)}}{4},\pm\frac{\sqrt{D_+(\gamma)}-\sqrt{D_-(\gamma)}}{4}\right\}.\label{eq:L4-exact-spectrum}
\end{eqnarray}
This expression is the exact solution used for the \(L=4\) comparison in Figure~\ref{figa}.

For every vector listed below, the eigenvalue equation is written directly as $H_\gamma v_E=E\, v_E$.
Since this spin-chain matrix is complex symmetric, the corresponding left eigenvector can be chosen as \(v_E^T\), up to normalization. The vectors below are not normalized. At isolated values where a displayed denominator vanishes, the corresponding eigenvector is understood by taking the appropriate limiting form or by choosing an equivalent normalization.

The four eigenvalue branches are
\begin{eqnarray}
    E=\frac{1}{2}, -\frac{1}{2},\frac{\gamma}{2}, -\frac{\gamma}{2}.
\end{eqnarray}
A corresponding set of eigenvectors is
\begin{eqnarray}
v_{1/2}&=&\left(0,0,0,-1,0,1,0,0,0,0,-1,0,1,0,0,0\right)^T,\\
v_{-1/2}&=&\left(0,0,0,-1,0,-1,0,0,0,0,1,0,1,0,0,0\right)^T,\\
v_{\gamma/2}&=&\left(-1,0,0,0,0,0,1,0,0,-1,0,0,0,0,0,1\right)^T,\\
v_{-\gamma/2}&=&\left(-1,0,0,0,0,0,-1,0,0,1,0,0,0,0,0,1\right)^T.
\end{eqnarray}
The remaining twelve eigenvectors are most compactly written in families. First, define
\begin{equation}
    E^{(1)}_{\pm}=\frac{\gamma-1\pm\sqrt{D_-(\gamma)}}{4} \qquad \mathrm{and} \qquad 
    a^{(1)}_{\pm}=\frac{\gamma-2E^{(1)}_{\pm}}{\gamma-1},\qquad b^{(1)}_{\pm}=-\frac{\gamma-2E^{(1)}_{\pm}}{\gamma-1}.
\end{equation}
Then
\begin{eqnarray}
v^{(1)}_{\pm}&=&\left(0,-1,a^{(1)}_{\pm},0,a^{(1)}_{\pm},0,0,1,-1,0,0,b^{(1)}_{\pm},0,b^{(1)}_{\pm},1,0\right)^T,
\end{eqnarray}
with $H_\gamma v^{(1)}_{\pm}=E^{(1)}_{\pm}v^{(1)}_{\pm}$. Second, define
\begin{equation}
E^{(2)}_{\pm}=\frac{1+\gamma\pm\sqrt{D_+(\gamma)}}{4} \qquad \mathrm{and} \qquad 
a^{(2)}_{\pm}=\frac{\gamma-2E^{(2)}_{\pm}}{\gamma+1},\qquad b^{(2)}_{\pm}=-\frac{\gamma-2E^{(2)}_{\pm}}{\gamma+1}.
\end{equation}
Then
\begin{eqnarray}
v^{(2)}_{\pm}&=&\left(0,1,a^{(2)}_{\pm},0,b^{(2)}_{\pm},0,0,-1,-1,0,0,b^{(2)}_{\pm},0,a^{(2)}_{\pm},1,0\right)^T,
\end{eqnarray}
with $H_\gamma v^{(2)}_{\pm}=E^{(2)}_{\pm}v^{(2)}_{\pm}$. 
Third, define
\begin{equation}
E^{(3)}_{\pm}=\frac{1-\gamma\pm\sqrt{D_-(\gamma)}}{4} \qquad \mathrm{and} \qquad 
a^{(3)}_{\pm}=\frac{-\gamma-2E^{(3)}_{\pm}}{\gamma-1},\qquad b^{(3)}_{\pm}=\frac{\gamma+2E^{(3)}_{\pm}}{\gamma-1}.
\end{equation}
Then
\begin{eqnarray}
v^{(3)}_{\pm}&=&\left(0,-1,a^{(3)}_{\pm},0,b^{(3)}_{\pm},0,0,-1,1,0,0,a^{(3)}_{\pm},0,b^{(3)}_{\pm},1,0\right)^T,
\end{eqnarray}
with $H_\gamma v^{(3)}_{\pm}=E^{(3)}_{\pm}v^{(3)}_{\pm}$. 
Fourth, define
\begin{equation}
    E^{(4)}_{\pm}=-\frac{1+\gamma\pm\sqrt{D_+(\gamma)}}{4} \qquad \mathrm{and} \qquad 
    a^{(4)}_{\pm}=\frac{-\gamma-2E^{(4)}_{\pm}}{\gamma+1}.
\end{equation}
Then
\begin{eqnarray}
v^{(4)}_{\pm}=\left(0,1,a^{(4)}_{\pm},0,a^{(4)}_{\pm},0,0,1,1,0,0,a^{(4)}_{\pm},0,a^{(4)}_{\pm},1,0\right)^T,
\end{eqnarray}
with $H_\gamma v^{(4)}_{\pm}=E^{(4)}_{\pm}v^{(4)}_{\pm}$. 
The final four eigenvectors correspond to the four choices
\begin{eqnarray}
    E^{(5)}_{\eta,s}&=&-\frac{\eta\sqrt{D_+(\gamma)}+s\sqrt{D_-(\gamma)}}{4},\qquad \eta,s=\pm1 .
\end{eqnarray}
For each such \(E^{(5)}_{\eta,s}\), define
\begin{eqnarray}
    c_{\eta,s}&=&\frac{4\left(E^{(5)}_{\eta,s}\right)^3-\left(5\gamma^2+8\right)E^{(5)}_{\eta,s}}{6\gamma},\\
    d_{\eta,s}&=&-\frac{5\gamma}{4}+\frac{\left(E^{(5)}_{\eta,s}\right)^2}{\gamma},\\
    e_{\eta,s}&=&\frac{-4\left(E^{(5)}_{\eta,s}\right)^3+\left(5\gamma^2+2\right)E^{(5)}_{\eta,s}}{3\gamma}.
\end{eqnarray}
Then
\begin{eqnarray}
v^{(5)}_{\eta,s}&=&\left(1,0,0,c_{\eta,s},0,d_{\eta,s},e_{\eta,s},0,0,e_{\eta,s},d_{\eta,s},0,c_{\eta,s},0,0,1\right)^T,
\end{eqnarray}
with $H_\gamma v^{(5)}_{\eta,s}=E^{(5)}_{\eta,s}v^{(5)}_{\eta,s}$. 

This explicit solution also displays the \(L=4\) EP structure directly. 
For the EP used in Fig.~\ref{figa}, $\gamma_{\rm EP}=\frac{3}{5}+\frac{4}{5}i$, 
%\begin{eqnarray}
%    \gamma_{\rm EP}&=&\frac{3}{5}+\frac{4}{5}i,
%\end{eqnarray}
one finds \(\sqrt{D_-(\gamma_{\rm EP})}=0\), while \(\sqrt{D_+(\gamma_{\rm EP})}\) remains finite. The eigenvalue pairs containing \(\sqrt{D_-(\gamma)}\) become degenerate at \(\gamma_{\rm EP}\):
%\begin{eqnarray}
%    E^{(1)}_{+}&=&E^{(1)}_{-},\\
%    E^{(3)}_{+}&=&E^{(3)}_{-},\\
%    E^{(5)}_{\eta,+}&=&E^{(5)}_{\eta,-},\qquad \eta=\pm1 .
%\end{eqnarray}
\begin{equation}
    E^{(1)}_{+}=E^{(1)}_{-}, \qquad 
    E^{(3)}_{+}=E^{(3)}_{-}, \qquad 
    E^{(5)}_{\eta,+}=E^{(5)}_{\eta,-},\quad \eta=\pm1 \,.
\end{equation}
The corresponding right eigenvectors reduce to the same vector:
%\begin{eqnarray}
%    v^{(1)}_{+}&=&v^{(1)}_{-},\\
%    v^{(3)}_{+}&=&v^{(3)}_{-},\\
%    v^{(5)}_{\eta,+}&=&v^{(5)}_{\eta,-},\qquad \eta=\pm1 .
%\end{eqnarray}
\begin{equation}
    v^{(1)}_{+}=v^{(1)}_{-},\qquad
    v^{(3)}_{+}=v^{(3)}_{-},\qquad
    v^{(5)}_{\eta,+}=v^{(5)}_{\eta,-},\quad \eta=\pm1 \,.
\end{equation}
This gives four degenerate pairs of many-body eigenvectors, consistent with the counting \(2^{L-2}=4\) for \(L=4\). Their distinct energies at this EP are
\begin{equation}
    E\in\left\{\pm\frac{-1+2\mathrm{i}}{10},
    \ \pm\frac{(2+\mathrm{i})\sqrt{15}}{10}\right\}.
\end{equation}
Direct row reduction of the \(16\times16\) spin Hamiltonian at \(\gamma=(3+4\mathrm{i})/5\) gives, for each of these four energies,
\begin{equation}
    \operatorname{rank}(H_\gamma-EI)=15,\qquad
    \operatorname{rank}\bigl[(H_\gamma-EI)^2\bigr]=14.
\end{equation}
Together with their algebraic multiplicity two in Eq.~\eqref{eq:L4-exact-spectrum}, these ranks establish one Jordan block of size two at each energy. The remaining eight eigenvalues are simple, so there are twelve independent eigenvectors at the EP, consistently with the counting in the main text.
Since the left eigenvector may be chosen as the transpose of the right eigenvector in this complex-symmetric representation, the left-right overlap is obtained from \(v_E^T v_E\), up to normalization. At the EP the degenerate eigenvectors are self-orthogonal, giving the zero of \(\langle L|R\rangle\) shown in Fig.~\ref{figa}.

For the overlap figures, the vectors \(v^{(5)}_{+,s}\) above fix the phase by their first component \(1\). Dividing by their positive Euclidean norm gives Eq.~\eqref{eq:overlap-normalization}. Explicitly,
\begin{equation}
    \rho_s=
    \frac{1+c_{+,s}^2+d_{+,s}^2+e_{+,s}^2}
    {1+|c_{+,s}|^2+|d_{+,s}|^2+|e_{+,s}|^2}.
\end{equation}
At the EP this numerator vanishes while the denominator remains positive. This is the zero displayed in Figure~\ref{figa}, with a finite Euclidean-normalized eigenvector.

\section{List of exceptional points for the non-Hermitian XY model}\label{EPsList}
In this Appendix, we list the locations of EPs with increasing of system sizes. The values are obtained by using Eqs.~(\ref{rooteqI}) and (\ref{rooteqII}) in the main text together with the double-root conditions in Eq.~(\ref{eq:double-root-condition}).
\begin{table}[H]
    \centering
    \setlength{\tabcolsep}{6pt}
    \renewcommand{\arraystretch}{1.15}
    \begin{tabular}{ccc}
        \toprule
        \(L\) & \(\gamma_{\mathrm{EP}}^{\mathrm{I}}\) & \(\gamma_{\mathrm{EP}}^{\mathrm{II}}\) \\
        \midrule
        4  & \(-0.6000 \pm 0.8000\, \mathrm{i} \) & \(0.6000 \pm 0.8000\, \mathrm{i} \) \\
        \midrule
        6  & \(-0.8030 \pm 1.3107\, \mathrm{i} \) & \(0.8030 \pm 1.3107\, \mathrm{i} \) \\
        6  & \(-0.3399 \pm 0.5547\, \mathrm{i} \) & \(0.3399 \pm 0.5547\, \mathrm{i} \) \\
        \midrule
        8  & \(-1.0116 \pm 1.7804\, \mathrm{i} \) & \(1.0116 \pm 1.7804\, \mathrm{i} \) \\
        8  & \(-0.4138 \pm 0.9104\, \mathrm{i} \) & \(0.4138 \pm 0.9104\, \mathrm{i} \) \\
        8  & \(-0.2413 \pm 0.4246\, \mathrm{i} \) & \(0.2413 \pm 0.4246\, \mathrm{i} \) \\
        \midrule
        10 & \(-1.2233 \pm 2.2336\, \mathrm{i} \) & \(1.2233 \pm 2.2336\, \mathrm{i} \) \\
        10 & \(-0.4893 \pm 1.2264\, \mathrm{i} \) & \(0.4893 \pm 1.2264\, \mathrm{i} \) \\
        10 & \(-0.2806 \pm 0.7035\, \mathrm{i} \) & \(0.2806 \pm 0.7035\, \mathrm{i} \) \\
        10 & \(-0.1886 \pm 0.3444\, \mathrm{i} \) & \(0.1886 \pm 0.3444\, \mathrm{i} \) \\
        \midrule
        12 & \(-1.4367 \pm 2.6784\, \mathrm{i} \) & \(1.4367 \pm 2.6784\, \mathrm{i} \) \\
        12 & \(-0.5666 \pm 1.5242\, \mathrm{i} \) & \(0.5666 \pm 1.5242\, \mathrm{i} \) \\
        12 & \(-0.3192 \pm 0.9477\, \mathrm{i} \) & \(0.3192 \pm 0.9477\, \mathrm{i} \) \\
        12 & \(-0.2143 \pm 0.5764\, \mathrm{i} \) & \(0.2143 \pm 0.5764\, \mathrm{i} \) \\
        12 & \(-0.1555 \pm 0.2899\, \mathrm{i} \) & \(0.1555 \pm 0.2899\, \mathrm{i} \) \\
        \midrule
        14 & \(-1.6512 \pm 3.1183\, \mathrm{i} \) & \(1.6512 \pm 3.1183\, \mathrm{i} \) \\
        14 & \(-0.6452 \pm 1.8120\, \mathrm{i} \) & \(0.6452 \pm 1.8120\, \mathrm{i} \) \\
        14 & \(-0.3587 \pm 1.1746\, \mathrm{i} \) & \(0.3587 \pm 1.1746\, \mathrm{i} \) \\
        14 & \(-0.2378 \pm 0.7787\, \mathrm{i} \) & \(0.2378 \pm 0.7787\, \mathrm{i} \) \\
        14 & \(-0.1744 \pm 0.4898\, \mathrm{i} \) & \(0.1744 \pm 0.4898\, \mathrm{i} \) \\
        14 & \(-0.1326 \pm 0.2505\, \mathrm{i} \) & \(0.1326 \pm 0.2505\, \mathrm{i} \) \\
        \bottomrule
    \end{tabular}
    \caption{Exceptional points in non-Hermitian XY model for different system sizes \(L\).}
    \label{tab:ep_values}
\end{table}

\bibliography{}

\end{document}